\documentclass[usenatbib]{mnras}
\usepackage{makecell}
\usepackage{graphicx}
\usepackage{dcolumn}
\usepackage{amssymb} 
\usepackage{amsmath}
\usepackage{ctable}
\usepackage{float}
\usepackage{bm}
\usepackage{float}
\usepackage{epsfig}
\usepackage{epstopdf}
\usepackage{epsf,color}
\usepackage{comment}
\usepackage{appendix}
\usepackage{aas_macros}
\usepackage{url}
\usepackage{soul}
\usepackage{widetext}
\usepackage{color, colortbl}
\def\bmsign{\mathchar"017E}
\def\dvecsign{\smash{\stackon[-1.95pt]{\bmsign}{\rotatebox{180}{$\bmsign$}}}}
\def\dvec#1{\def\useanchorwidth{T}\stackon[-4.2pt]{#1}{\,\dvecsign}}
\usepackage{stackengine}
\stackMath
\newcommand{\cmnt}[1]{}

\usepackage[normalem]{ulem}
\usepackage{color}

\usepackage{tgtermes}

\usepackage{mathtools}

\title[\textsc{Galacticus} calibration]{A new calibration method of sub-halo orbital evolution for semi-analytic models}

\author[S. Yang et al.]{
Shengqi Yang,$^{1}$\thanks{E-mail:sy1823@nyu.edu}
Xiaolong Du,$^{2}$\thanks{E-mail:xdu@carnegiescience.edu}
Andrew J. Benson, $^{2}$
Anthony R. Pullen,$^{1,3}$
\newauthor
Annika H. G. Peter$^{4,5}$\\
$^{1}$Center for Cosmology and Particle Physics, Department of Physics, New York University, 726 Broadway, New York, NY, 10003, U.S.A.\\
$^{2}$Carnegie Observatories, 813 Santa Barbara Street, Pasadena, CA 91101, U.S.A\\
$^{3}$Center for Computational Astrophysics, Flatiron Institute, New York, NY 10010, U.S.A.\\
$^{4}$CCAPP and Department of Physics, The Ohio State University, 191 W. Woodruff Ave., Columbus, OH 43210, USA\\
$^{5}$Department of Astronomy, The Ohio State University, 140 W. 18th Ave., Columbus, OH 43210, USA
}

\date{Accepted XXX. Received YYY; in original form ZZZ}

\pubyear{2020}

\begin{document}
\label{firstpage}
\pagerange{\pageref{firstpage}--\pageref{lastpage}}
\maketitle

\begin{abstract}
Understanding the non-linear dynamics of satellite halos (a.k.a. ``sub-halos'') is important for predicting the abundance and distribution of dark matter substructures and satellite galaxies, and for distinguishing among microphysical dark matter models using observations. Typically, modeling these dynamics requires large N-body simulations with high resolution. Semi-analytic models can provide a more efficient way to describe the key physical processes such as dynamical friction, tidal mass loss, and tidal heating, with only a few free parameters. In this work, we present a fast Monte Carlo Markov Chain fitting approach to explore the parameter space of such a sub-halo non-linear evolution model. We use the dynamical models described in an earlier work and calibrate the models to two sets of high-resolution cold dark matter N-body simulations, ELVIS and Caterpillar. Compared to previous calibrations that used manual parameter tuning, our approach provides a more robust way to determine the best-fit parameters and their posterior probabilities. We find that jointly fitting for the sub-halo mass and maximum velocity functions can break the degeneracy between tidal stripping and tidal heating parameters, as well as providing better constraints on the strength of dynamical friction. We show that our semi-analytic simulation can accurately reproduce N-body simulations statistics, and that the calibration results for the two sets of N-body simulations agree at 95\% confidence level. Dynamical models calibrated in this work will be important for future dark matter substructure studies.
\end{abstract}




\section{Introduction}
Exploring the physics behind galaxy and star formation is one of the major concerns of modern astrophysics. The simple cold dark matter (CDM) paradigm successfully explains large-scale cosmic properties, including the cosmic microwave background \citep{1982ApJ...263L...1P} and the large-scale structure (LSS) of galaxy distributions \citep{2011A&A...536A...1P,2014A&A...571A...1P,2012MNRAS.427.3435A}. However, on galactic scales, several puzzles such as the core-vs.-cusp problem \citep{1980ApJ...238..471R,1981AJ.....86.1825B,1988MNRAS.234..131P,1996MNRAS.281...27P,2001MNRAS.320L...1S,2004MNRAS.353L..17D,2009MNRAS.397.1169D,2009ApJ...706.1078N,2011ApJ...728L..39N,2010AdAst2010E...5D,2011ApJ...741L..29K,2011MNRAS.414.3617K,2012MNRAS.420.2034S,2012arXiv1203.4240W,2019ApJ...873....5R,2019ApJ...887...94R} and the missing satellite problem \citep{1993MNRAS.264..201K,1999ApJ...522...82K,2010arXiv1009.4505B,2011MNRAS.415L..40B,2012MNRAS.422.1203B,2012MNRAS.424.2715W} still remain to be fully explained. Many possible solutions, including baryonic feedback \citep{2007arXiv0704.3078M,2014ApJ...786...87B,2018PhRvL.121u1302K}, and modified dark matter (DM) models \citep{2004ApJ...606..819M,2005A&A...438..419B,2005MNRAS.363.1092A,2008ApJ...679.1173R,2012MNRAS.420.2318L,2016PhRvL.116d1302K}, have been proposed and tested via N-body and hydrodynamical simulations \citep{2017ARA&A..55..343B,robles2017,2019MNRAS.483.4086B,2020MNRAS.tmp.2452L}, although whether any of the proposed models can fully explain the deviation of CDM expectations from observational results remains unclear. A variety of upcoming experimental measurements \citep{2005ApJ...621..757S,2009MNRAS.399L..39V}, especially future strong lensing surveys \citep{2009ApJ...699.1720K,2009MNRAS.400.1583V,2010MNRAS.407..225V,2012Natur.481..341V,2018MNRAS.481.3661V,2016JCAP...11..048H,2017JCAP...05..037B,2018MNRAS.478.4816S,2019MNRAS.487.5721G,2020MNRAS.491.6077G,2019ApJ...883...14M,2020MNRAS.492.3047H,2020MNRAS.492.5314N} and studies of the stellar halo of the Milky Way \citep{2011ApJ...731...58Y,2014ApJ...788..181N,2015ApJ...803...75N,2016MNRAS.463..102E,2017MNRAS.466..628B,2018PhRvL.120u1101B,2018JCAP...07..061B,2019arXiv191102663B,2018ApJ...867..101B,2020ApJ...892L..37B,2018JCAP...07..041V,2019ApJ...884...51G,2019ApJ...872..152I,2019PhRvL.123i1101R,2020arXiv200201938M,2020PhRvD.102b3026M}, will be able to probe small-scale DM structures with high resolution.\par 
To constrain DM properties with future observational results, rapid and accurate simulations are needed to provide theoretical predictions. One approach to achieve fast and physically grounded simulations is to use semi-analytic models (SAMs). Instead of solving the differential equations that describe the motion of each N-body particle, SAMs approximate the merging history of a DM halo using the extended Press-Schechter (EPS) formalism \citep{1974ApJ...187..425P,1991ApJ...379..440B,1991MNRAS.248..332B,1993MNRAS.262..627L,parkinson_generating_2008}. SAMs also replace computationally expensive hydrodynamic simulations by simplified but physically motivated treatments of gas cooling, star formation, stellar feedback, and galaxy merging. As an intermediate approach between analytic theory and N-body simulations, SAMs are transparent about the underlying assumptions and are computationally efficient in exploring the large parameter space of unknown physical processes. One free and open source SAM---\textsc{Galacticus}---is developed by \cite{2012NewA...17..175B}. The key feature of \textsc{Galacticus} is its modularity---different models that describe identical physical process can be added and compared easily. \par 
The EPS formalism used by SAMs has been calibrated to CDM N-Body simulations. \cite{2013MNRAS.428.1774B} generalized the EPS formalism implemented in \textsc{Galaticus} to the warm dark matter (WDM) model. This enables \textsc{Galacticus} to self-consistently predict the variation of DM statistics under the CDM and WDM paradigms. \citeauthor{2014ApJ...792...24P}~(\citeyear{2014ApJ...792...24P}; here after AP2014) then added models that describe the orbital evolution and mass loss of sub-halos within host halos by accounting for dynamical friction, tidal stripping, and tidal heating, and studied how these non-linear effects influence the sub-halo distribution under the CDM and WDM paradigms. Specifically, AP2014 adopted the dynamical friction Coulomb logarithm proposed by \cite{2001ApJ...559..716T} and the tidal heating adiabatic index proposed by \cite{1999ApJ...513..626G}. The tidal effect models were then calibrated to the Aquarius CDM N-body simulation \citep{2008MNRAS.391.1685S} through manual parameter tuning. The dynamical friction model and the calibrated tidal effect models were then applied to WDM halos. AP2014 showed qualitatively that the sub-halo mass function is a useful tracer of sub-halo-host interactions, and provided evidence that DM halo statistical properties such as the sub-halo mass function and density profiles differ between CDM and WDM models when the sub-halo non-linear evolution mechanisms are fixed. These findings point to the potential of using sub-halo statistics to differentiate DM microphysics. Well-calibrated sub-halo non-linear evolution models, which are necessary for generating accurate DM substructure statistical predictions, are therefore important for DM property constraints. However, AP2014 did not vary the Coulomb logarithm for dynamical friction, nor the adiabatic index for tidal heating. A full search of the parameter space through a Monte Carlo Markov Chain (MCMC) fit was also not performed. Therefore, reliable and accurate values of model parameters applicable for future studies are still 
not well quantified.\par 
  
In this work, we introduce an MCMC fitting workflow to fully explore the parameter space with high efficiency. We apply this MCMC fitting method to calibrate the dynamical friction, tidal stripping, and tidal heating models introduced in AP2014 to the ELVIS \citep{2014MNRAS.438.2578G} and Caterpillar \citep{2016ApJ...818...10G} CDM N-body simulations of Milky Way-sized host halos. Non-linear sub-halo evolution models calibrated in this work can provide more
accurate semi-analytical predictions for sub-halo statistics and place more robust constraints on DM
microphysics with future observations. This MCMC fitting workflow is also applicable for non-linear evolution model refinements in the future.\par
The plan of this paper is as follows. In Section~\ref{sec:2} we review the dynamical friction, tidal stripping, and tidal heating models implemented in \textsc{Galacticus}. In Section \ref{sec:3} we introduce ELVIS and Caterpillar----the two sets of Milky Way-sized N-body simulations we use in this work. We also present relevant parameter settings in the corresponding \textsc{Galacticus} simulations. We introduce our fast MCMC fitting strategy as well as the fitting results in Section \ref{sec:4}. We discuss the physical meaning behind the MCMC results in Section \ref{sec:5} and conclude in Section \ref{sec:6}.\par

\section{non-linear evolution theory}\label{sec:2}
\begin{figure}
    \centering
    \includegraphics[width=0.45\textwidth]{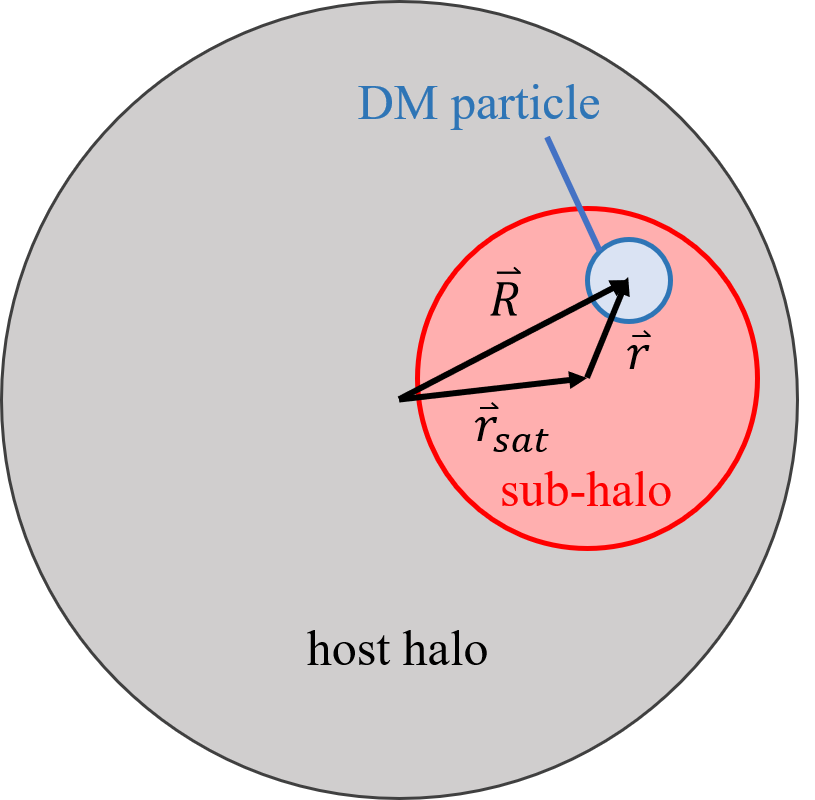}
    \caption{Geometry of a simplified host-satellite-DM particle system used in the non-linear evolution theory. The grey circle represents the host halo, the red circle is a sub-halo, and the blue circle is a DM particle member of the sub-halo.   $\vec{r}_{\mathrm{sat}}$ is the position vector pointing from the host to the satellite halo. $\vec{R}$ is the relative position from the host to the DM particle member. $\vec{r}$ is the relative position from the satellite center to its DM particle.}\label{fig:1}
\end{figure}

In this section we give a brief review of the models for three key, non-linear evolution processes---dynamical friction, tidal stripping, and tidal heating---implemented in \textsc{Galacticus} by AP2014. The geometry of a simplified system which consists of a host halo, a satellite, and a DM particle of the satellite is presented in Figure~\ref{fig:1} to clarify different position vectors involved in the non-linear evolution models. We also refer readers to \cite{2001ApJ...559..716T,2002MNRAS.333..156B,2005ApJ...624..505Z} for further details.\par
The DM halo evolution engine in \textsc{Galacticus} works as follows. First, merger trees are constructed (using the EPS formalism, specifically the algorithm proposed by \citealt{parkinson_generating_2008}) backward in time until the required mass resolution is reached along each branch. The properties of halos are then evolved forward in time. When two halos encounter each other in a merger tree, the more massive becomes the host with the less massive one becoming a satellite (sub-halo) within that host. The satellite is initially placed isotropically at random on the sphere corresponding to the virial radius of the host, and is given an initial velocity drawn from a distribution obtained from cosmological simulations, with the radial component directed inward, and the direction of the tangential component sampled isotropically at random. Variation of the satellite initial velocity distribution can significantly influence the simulated sub-halo statistics. In this work we choose the velocity distribution measured by \cite{2015MNRAS.448.1674J}, which is based on the DOVE $\Lambda$CDM N-Body simulation with mass resolution $2\times 10^8 \mathrm{M}_\odot$. We present the influence of the satellite initial velocity distribution on the model calibration results in more detail in Appendix \ref{apdx:1}. The position within the host, bound mass, and density profile of the satellite are then tracked until certain merging/disruption criteria are satisfied at which point the satellite is considered to be completely disrupted (merged with the host) and is removed.\par
Several assumptions are made in \textsc{Galacticus} to achieve fast simulation. As \textsc{Galacticus} dynamically evolves the positions and velocities of a satellite, the masses of other satellites are treated as a part of the host halo and the detailed sub-halo--sub-halo interactions are ignored. \cite{2005MNRAS.364..977P} shows that such interactions have negligible influence on the mass and spatial distribution of the substructures. In this work, \textsc{Galacticus} classifies a satellite as being destroyed by its host if: 1) the distance between the sub-halo and the host halo is smaller than a fraction $f$ of the host virial radius; or 2) the sub-halo mass falls below a specified mass resolution $M_{\mathrm{res}}$. These criteria are adjustable in \textsc{Galacticus} and can be changed for different applications. In this work we take $f=0.01$ and $M_{\mathrm{res}}=5\times10^7\mathrm{M}_\odot$. We have checked that these two criteria are sufficient for the sub-halo mass range we consider in this work. Further details about the \textsc{Galacticus} mass resolution settings are presented in Section \ref{sec:3}.\par

\subsection{Dynamical Friction}
We assume that as a DM sub-halo with mass $M$ and velocity $\bm{V}_{\mathrm{sat}}$ travels through the sea of host halo DM particles, the sub-halo will experience a steady deceleration, known as dynamical friction. Dynamical friction arises as the sub-halo deflects nearby DM particles through gravitational interaction, and thus creates an overdense region behind it. This accelerates the sub-halo opposite to its direction of motion, slowing it down. First proposed by \cite{1943RvMP...15....1C} to describe the motion of a body through a uniform medium, the dynamical friction equation can be applied to bodies traveling through finite media with only minor modification \citep{1986ApJ...300...93W}. If we assume that the distribution of host particles is reasonably well modeled by a Maxwell-Boltzmann distribution \citep{1996APh.....6...87L, 2013ApJ...764...35M}, the Chandrasekhar formula gives the acceleration of the sub-halo caused by dynamical friction $\bm{a}_{\mathrm{df}}$ as:
\begin{equation}
\begin{split}
    \bm{a}_{\mathrm{df}}=&-4\pi G^2 \ln\Lambda M_{\mathrm{sat}}\rho_{\mathrm{host}}(r_{\mathrm{sat}})\dfrac{\bm{V}_{\mathrm{sat}}}{V_{\mathrm{sat}}^3}\\
    &\times\left[{\rm erf}(X_v)-\dfrac{2X_v}{\sqrt{\pi}}\exp(-X_v^2)\right]\,,
\end{split}\label{eq:1}
\end{equation}
where $r_{\mathrm{sat}}$ is the sub-halo position within the host, $X_v=V_{\mathrm{sat}} / \sqrt{2}\sigma_v$ with $\sigma_v$ the velocity dispersion of DM particles in the host. The range of impact parameters that contribute to the satellite deceleration is not well defined, and the Coulomb logarithm $\ln\Lambda$ is introduced to absorb the uncertainty of an effective impact parameter integration range. We treat $\ln\Lambda$ as a free parameter. We assume the host halo has an NFW density profile $\rho_{\mathrm{host}}$ \citep{1997ApJ...490..493N}:
\begin{equation}
    \rho_{\mathrm{host}}(r_{\mathrm{sat}})\propto\left(\dfrac{r_{\mathrm{sat}}}{R_\mathrm{s}}\right)^{-1}\left(1+\frac{r_{\mathrm{sat}}}{R_\mathrm{s}}\right)^{-2}\,,
\end{equation}
where $R_\mathrm{s}$ is the scale length. The NFW profile is normalized such that the total halo mass is enclosed within the virial radius $R_{\mathrm{vir}}$. The halo concentration parameter $c \equiv R_{\mathrm{vir}}/R_\mathrm{s}$ is computed following \cite{2015ApJ...799..108D}.\par 
We use equation~(14) of \cite{2001MNRAS.321..155L} to calculate $\sigma_v(r_\mathrm{sat})$, which is better physically-motivated than the approach used in AP2014, where $\sigma_v$ is approximated by the virial velocity of the host halo $V_{\mathrm{vir}}$.

\subsection{Tidal Stripping}
While the satellite orbits its host, it is subjected to tidal forces, which pull the satellite material on the near side toward the host center and in the opposite direction on the far side. When the tidal force is larger than the gravitational force from the satellite itself, material in the satellite could become unbound, forming tidal tails. The radius at which the tidal force equals the self-gravity force is called the tidal radius. To first order, the tidal force is proportional to the gradient of gravitational force from the host at the satellite position and to the distance from the satellite center. Thus, the satellite will be stripped outside-in as the pericenter of its orbit moves ever closer to the host center due to dynamical friction, and as the sub-halo's density drops due to tidal heating. A summary of various definitions of tidal radius is presented in \cite{2018MNRAS.474.3043V}. Taking into account the extended sub-halo mass profile and the motion of particles within the satellite, \textsc{Galacticus} computes the tidal radius, $r_t$, as \citep{1962AJ.....67..471K,2018MNRAS.474.3043V}:

\begin{equation}
    r_t=\left(\dfrac{GM_{\mathrm{sat}}(<r_t)}{\omega^2-\left.\frac{\mathrm{d}^2\Phi}{\mathrm{d}R^2}\right\vert_{r_{\mathrm{sat}}}}\right)^{1/3}.
\end{equation}
 Here $M_{\mathrm{sat}}(<r_t)$ is the satellite mass enclosed within the tidal radius, $\omega$ is the angular frequency of the satellite orbit, and $R$ is the distance from the center of the host halo to the satellite DM particle. Here we have assumed that the satellite and its DM particles are orbiting within the host with a common angular frequency. Since we assume a spherically symmetric NFW profile, $\rho_{\mathrm{host}}$, for the host halo, the second derivative of the gravitational potential from the host $\mathrm{d}^2\Phi/\mathrm{d}R^2$ is given by:
\begin{equation}
    \left.\dfrac{\mathrm{d}^2\Phi}{\mathrm{d}R^2}\right\vert_{r_{\mathrm{sat}}}=-\dfrac{2GM(<r_{\mathrm{sat}})}{r^3_{\mathrm{sat}}}+4\pi G\rho_{\mathrm{halo}}(r_{\mathrm{sat}})\,.
\end{equation}
Following \cite{2005ApJ...624..505Z}, \textsc{Galacticus} models the tidal stripping effect by assuming that the satellite mass outside $r_t$ is lost on an orbital time scale:
\begin{equation}
    \dfrac{\mathrm{d}M_{\mathrm{sat}}}{\mathrm{d}t}=-\alpha\dfrac{M_{\mathrm{sat}}(>r_t)}{T_{\mathrm{orb}}}\,.
\end{equation}
Here we define the instantaneous orbital period as the minimum of the instantaneous angular and radial periods $T_{\mathrm{orb}}=\min(2\pi/\omega,2\pi r_{\mathrm{sat}}/V_{\mathrm{sat}})$, and $\alpha$ is treated as a free parameter.\par

\subsection{Tidal Heating}
The host halo not only strips mass from the satellite through gravitational tides, but also introduces an additional velocity dispersion to the satellite particles. The extra random motion within the satellite caused by the rapidly varying tidal field heats up the satellite. As a result, tidal heating will cause the satellite to expand and a larger fraction of the satellite mass will extend outside the tidal radius and become subjected to tidal stripping. \par
\textsc{Galacticus} models tidal heating following \cite{1997astro.ph..9161G} and \cite{2001ApJ...559..716T}. Under the impulse approximation, the heating rate introduced by this effect, averaged over all the randomly distributed DM particle members can be modeled as \citep{2001ApJ...559..716T}:

\begin{equation}\label{eq:13}
\begin{split}
    \left\langle\dfrac{\mathrm{d}E}{\mathrm{d}t}\right\rangle=\dfrac{1}{3}r^2(t)g_{ab}(t)G_{ab}(t).
\end{split}
\end{equation}
Here $r$ is the distance between the satellite center and the DM particle, $g$ is the tidal tensor, and $G$ is the time integral of $g$:
\begin{equation}\label{eq:tidal_int}
G_{ab}=\int_0^t\mathrm{d}t'\left[g_{ab}(t')-G_{ab}(t')/T_{\mathrm{orb}}\right].
\end{equation}
Here we have added a decaying term $-G_{ab}(t')/T_{\mathrm{orb}}$ in the integrand considering that the positions of DM particles have non-negligible changes in one satellite orbital time, thus the impulse approximation is not valid on time scales larger than $T_{\mathrm{orb}}$.\par
\cite{1999ApJ...513..626G} points out that although the tidal heating in the sub-halo outskirts is well described by the impulse approximation, the effect in the inner part (where dynamical times in the sub-halo may be comparable to the shock timescale) is more complex. These more strongly bound satellite particles respond more adiabatically to the tidal heating process, and the conservation of the adiabatic invariant suppresses the heating shock. On the other hand, resonances in the system will strengthen the effects of the shock. To account for the breakdown of the impulse approximation where the shock duration becomes comparable to the orbital time scale as well as the high order heating effects, AP2014 modify equation~(\ref{eq:13}) as:
\begin{equation}
\left\langle\dfrac{\mathrm{d}E}{\mathrm{d}t}\right\rangle=\dfrac{\epsilon_h}{3}\left[1+\left(\omega_p T_{\mathrm{shock}}\right)^2\right]^{-\gamma}r^2g_{ab}(t)G_{ab}(t).
\end{equation}
The bracketed factor is the adiabatic correction discussed in \cite{1999ApJ...513..626G}, $T_{\mathrm{shock}} = r_{\mathrm{sat}}/V_{\mathrm{sat}}$ is the shock time scale, $\omega_p$ is the angular frequency of particles at the half-mass radius of the satellite \footnote{Here we follow the same definition as in \cite{1999ApJ...513..626G}, while AP2014 takes the orbital frequency of the satellite around the host.}. The heating coefficient, $\epsilon_h$, which accounts for the higher-order heating effects, is treated as a free parameter. AP2014 sets the adiabatic index $\gamma=2.5$ following \cite{1999ApJ...513..626G}. However, it has been shown that when $T_{\mathrm{shock}} \gg 1/\omega_p$, the suppression from adiabatic correction is shallower with $\gamma$ approaching to $1.5$ \citep{1994AJ....108.1398W,1994AJ....108.1403W,1999ApJ...513..626G}. There is also evidence that ignoring the adiabatic correction does not have a significant influence on sub-halo statistics when applied to cosmological simulations \citep{2018MNRAS.474.3043V}. In our MCMC simulation, we consider two limiting cases, $\gamma=0$ and $\gamma=2.5$. We will present the MCMC fitting results for both $\gamma$ values later in Section~\ref{sec:4}.
Energy injected into the satellite through tidal heating will cause the density profile to change. Under the assumption that each mass shell within the satellite stays virialized, and that there is no shell-crossing, AP2014 show that the satellite density profile can be modified as:
\begin{equation}
\label{eq:heatedProf}
\begin{split}
    \rho_{\mathrm{sat}}(r_f)=&\left[1-\dfrac{2r_i^3Q(r_i)}{GM_{\mathrm{sat}}(<r_i)}\right]^4\left[1+\dfrac{4r_i^3Q(r_i)}{GM_{\mathrm{sat}}(<r_i)}\right.\\
    &\left.-\dfrac{8\pi x_i^6Q(r_i)}{GM_{\mathrm{sat}}^2(<r_i)}\rho_{\mathrm{sat}}(r_i)\right]^{-1}\rho_{\mathrm{sat}}(r_i).
\end{split}
\end{equation}
Here $r_i$ and $r_f$ are the initial and final radii of a mass shell, and $Q(r_i)=E(r_i)/r_i^2$.

\subsection{Statistics for model constraint}\label{sec:2.4}
The sub-halo mass function is sensitive to satellite mass loss caused by tidal stripping and is therefore widely used to constrain DM phenomenology and clustering properties \citep{2010PhRvD..82l3521P,2012PhRvD..85d3514W,2014MNRAS.442.2487K,2014PASA...31....6M}. In this work we not only calibrate the three nonlinear evolution models with the sub-halo mass function at redshift $z=0$, but also consider the statistics of the present-time maximum circular velocity. We define the sub-halo mass, $M$, as the sub-halo’s gravitationally bound mass at $z=0$. To minimize the amplitude of fluctuations in the sub-halo mass function caused by the variation of host halo mass, we use the ratio between sub-halo mass and host halo mass as the mass variable of the sub-halo mass function. The advantages of a joint fit to $\mathrm{d}N/\mathrm{d}\log(M/M_{\mathrm{host}})$ and $V_{\mathrm{max}}(M)$ are shown below.\par
The parameters $\ln\Lambda$, $\alpha$, and $\epsilon_h$ effectively control the strength of dynamical friction, tidal stripping, and tidal heating in our semi-analytic simulation. Increasing $\ln\Lambda$ while fixing $\alpha$ and $\epsilon_h$ leads to greater deceleration of DM sub-halos caused by dynamical friction, thus more satellites merge into the host and $\mathrm{d}N/\mathrm{d}\log(M/M_{\mathrm{host}})$ decreases over the entire mass range. Since $a_{\mathrm{df}}\propto M$, massive halos are more sensitive to dynamical friction, leading to a steeper slope of $\mathrm{d}N/\mathrm{d}\log(M/M_{\mathrm{host}})$ as $\ln\Lambda$ increases. The maximum circular velocity of a DM halo is computed from the halo density profile Eq. (\ref{eq:heatedProf}):
\begin{equation}\label{eq:VmaxHeated}
\begin{split}
   &V^2(r)=\dfrac{G\int_0^{r}4\pi r'^2 \rho_{\mathrm{sat}}(r') \mathrm{d}r'}{r},\\
   &\dfrac{\mathrm{d}V(r)}{\mathrm{d}r}\Bigg|_{r=r_{\mathrm{max}}}=0,
  \end{split}
\end{equation}
where $r_{\mathrm{max}}$ is the distance between the satellite halo and its DM particle where the circular velocity of the DM particle reaches its maximum. Before the satellite falls into the host, its density profile has not been deformed by the tidal effects and maintains an NFW profile. Therefore at infall time the maximum circular velocity $V_{\mathrm{max(infall)}}$ of the DM halo with an NFW profile has the analytical form:
\begin{equation}\label{eq:vmaxnfw}
    \begin{split}
        V_{\mathrm{max(infall)}}&=0.465\times\sqrt{\dfrac{GM_{\mathrm{(infall)}}}{R_{\mathrm{vir(infall})}}}\sqrt{\dfrac{c_{\mathrm{(infall)}}}{f(c_{\mathrm{(infall)}})}},\\
        f(x)&=\ln(1+x)-\dfrac{x}{1+x}.
    \end{split}
\end{equation}

\begin{figure*}
    \centering
    \includegraphics[width=1.0\textwidth]{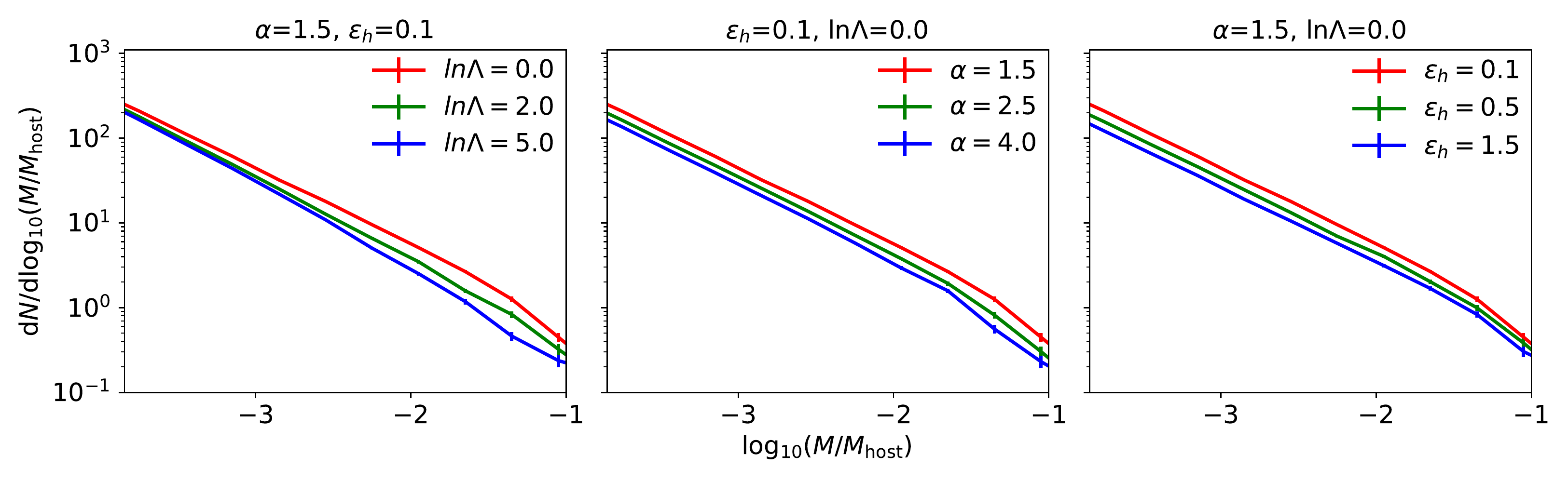}\\
    \includegraphics[width=1.0\textwidth]{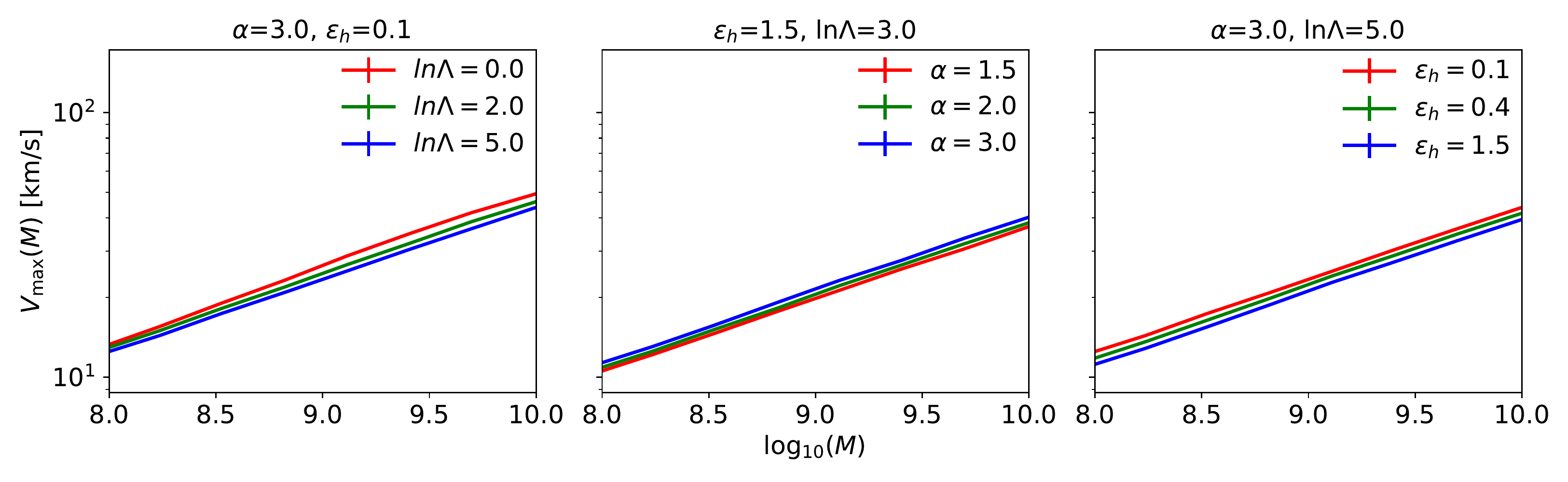}
    \caption{$\mathrm{d}N/\mathrm{d}\log(M/M_{\mathrm{host}})$ and $V_{\mathrm{max}}(M)$ simulated by \textsc{Galacticus} under different $\{\alpha,\epsilon_h,\ln\Lambda\}$ combinations at redshift $z=0$. \textsc{Galacticus} simulations are made with the Caterpillar cosmology, and setting the tidal heating parameter $\gamma=0$. Parameter combinations used in the plots are chosen such that the $\mathrm{d}N/\mathrm{d}\log(M/M_{\mathrm{host}})$ and $V_{\mathrm{max}}(M)$ changes are easy to see. Variations in $\alpha$ and $\epsilon_h$ influence the sub-halo mass function in the same direction, while $V_{\mathrm{max}}(M)$ varies in opposite directions. Therefore a joint fit for $\mathrm{d}N/\mathrm{d}\log(M/M_{\mathrm{host}})$ and $V_{\mathrm{max}}(M)$ can break the degeneracy between $\alpha$ and $\epsilon_h$. However, $\mathrm{d}N/\mathrm{d}\log(M/M_{\mathrm{host}})$ and $V_{\mathrm{max}}(M)$ change in the same direction as $\ln\Lambda$ and $\epsilon_h$ increase, thus we still see a negative correlation in the $\ln\Lambda-\epsilon_h$ contour in Figure~\ref{fig:6}. }\label{fig:2}
\end{figure*}

Here $M_{\mathrm{(infall)}}$, $c_{\mathrm{(infall)}}$, $R_{\mathrm{vir(infall)}}$ are the mass, concentration, and virial radius of the satellite when it first enters the host's virial radius. Equation~(\ref{eq:vmaxnfw}) shows that sub-halos with larger mass and concentration have larger $V_{\mathrm{max}}$---a statement that is true not only for NFW profile but also for general forms of $\rho_{\mathrm{sat}}$. Since sub-halos with large initial mass stay in the host for longer before they reach the disruption mass, and are more sensitive to dynamical friction, as $\ln\Lambda$ increases, the number of massive sub-halos with large $V_{\mathrm{max(infall)}}$ decreases, leading to a lower averaged $V_{\mathrm{max(infall)}}$ and a lower $V_{\mathrm{max}}$ at $z=0$ in the system. Semi-analytically simulated variations of $\mathrm{d}N/\mathrm{d}\log(M/M_{\mathrm{host}})$ and $V_{\mathrm{max}}(M)$ at $z=0$ caused by varying $\ln\Lambda$ are shown in the first column of Figure~\ref{fig:2}. \par

Increasing $\alpha$ while fixing $\epsilon_h$ and $\ln\Lambda$ corresponds to higher efficiency for the host halo to strip away satellite mass distributed outside of the tidal radius of the sub-halos, thus $\mathrm{d}N/\mathrm{d}\log(M/M_{\mathrm{host}})$ decreases over the entire mass range. However, the density profile of the satellites within the tidal radius is not influenced, such that a satellite with smaller mass can maintain its $V_{\mathrm{max}}$ under strong tidal stripping. As a result $V_{\mathrm{max}}(M)$ increases as $\alpha$ increases. The influences of $\alpha$ on $\mathrm{d}N/\mathrm{d}\log(M/M_{\mathrm{host}})$ and $V_{\mathrm{max}}(M)$ at $z=0$ are shown in the second column of Figure~\ref{fig:2}. \par

Finally, increasing $\epsilon_h$ while fixing $\alpha$ and $\ln\Lambda$ corresponds to stronger tidal heating. A larger fraction of mass within the satellite will extend beyond the tidal radius and so will be stripped by the tidal field of the host halo---this decreases $\mathrm{d}N/\mathrm{d}\log(M/M_{\mathrm{host}})$ over the entire mass range. Since the density profile of the satellite becomes less compact and a larger fraction of the satellite mass can be stripped off, $V_{\mathrm{max}}$ also decreases as $\epsilon_h$ increases. This phenomenon is presented in the third column of Figure~\ref{fig:2}.\par 
Notice that $\mathrm{d}N/\mathrm{d}\log(M/M_{\mathrm{host}})$ and $V_{\mathrm{max}}(M)$ vary differently as a result of increases in $\alpha$ and $\epsilon_h$. Thus a joint fit to $\mathrm{d}N/\mathrm{d}\log(M/M_{\mathrm{host}})$ and $V_{\mathrm{max}}(M)$ can break the degeneracy between $\alpha$ and $\epsilon_h$. However, $\mathrm{d}N/\mathrm{d}\log(M/M_{\mathrm{host}})$ and $V_{\mathrm{max}}(M)$ vary in similar ways with increases in $\epsilon_h$ and $\ln\Lambda$, thus we expect to see the negative correlation in the posterior distribution of $\epsilon_h$ and $\ln\Lambda$. Although $\epsilon_h$ only influences the amplitude of the sub-halo mass function while $\ln\Lambda$ also changes its slope, the limited size of the ELVIS and Caterpillar N-body simulations we use in this work mean that there are too few of the most massive satellites to fully break the $\epsilon_h-\ln\Lambda$ degeneracy. We expect this to also lead to a weak constraint on $\ln\Lambda$.\par

\section{N-body simulation and \textsc{Galacticus} settings}\label{sec:3}
In this work, we calibrate the three free parameters introduced in the dynamical friction and tidal effect models in the last section to two independent CDM N-body simulations---ELVIS and Caterpillar. We choose to calibrate non-linear evolution models against these two Milky Way-sized simulations for several reasons. Firstly, since the Milky Way and its satellite system have been studied extensively, Milky Way-sized N-Body simulations with high resolution are rich and easy to access. Secondly, using Milky Way-sized N-Body simulations for model calibration allows a direct comparison between this work and AP2014. Lastly, ELVIS and Caterpillar provide 24 and 34 isolated halo catalogs respectively. These sample volumes are much larger than all the other current Milky Way-sized N-Body simulations and can effectively suppress the statistical uncertainties caused by the halo-to-halo scatter. When calibrating \textsc{Galacticus} to Caterpillar we use \textit{Planck} cosmological parameters, $\Omega_m=0.32$, $\Omega_\Lambda=0.68$, $\sigma_8=0.83$, $n_s=0.96$, and $h=0.6711$ \citep{2014A&A...571A..16P}, while for ELVIS we use cosmological parameters given by \textit{Wilkinson Microwave Anisotropy Probe 7} $\Omega_m=0.266$, $\Omega_\Lambda=0.734$, $\sigma_8=0.801$, $n_s=0.963$, and $h=0.71$ \citep{2011ApJS..192...16L}.\par 
As described in Sec.~\ref{sec:2.4}, we use the sub-halo mass function and maximum circular velocity functions at redshift $z=0$ from these simulations as the constraints on our model. We expect to constrain tidal mass loss and dynamical friction through the sub-halo mass function $\mathrm{d}N/\mathrm{d}\log M$. Since tidal heating effects will extend the density profile of satellites and decrease the maximum circular velocity of satellites, we use the maximum circular velocity function $V_{\mathrm{max}}(M)$ to constrain tidal heating. Although $\mathrm{d}N/\mathrm{d}\log M$ is self-similar for CDM, the amplitude of $\mathrm{d}N/\mathrm{d}\log M$ is sensitive to the host halo mass. Each host halo in the N-body simulation has a slightly different mass, and the host halo mass distributions for ELVIS and Caterpillar differ. Averaging $\mathrm{d}N/\mathrm{d}\ln M$ over all the simulated host halos will introduce uncertainties to the sub-halo mass function amplitude and will further influence the parameter fitting accuracy. In order to minimize the effects of the distribution of host halo masses, we compute and calibrate the number of satellites in fractional mass bins $\mathrm{d}N/\mathrm{d}\log(M/M_{\mathrm{host}})$ instead. The maximum circular velocity is directly determined by the satellite mass $M$ and is independent of the host halo mass $M_{\mathrm{host}}$, so we fit the AP2014 model to $V_{\mathrm{max}}(M)$ instead of $V_{\mathrm{max}}(M/M_{\mathrm{host}})$.\par 
In this work we include only satellites within the host halo's virial radius for the $\mathrm{d}N/\mathrm{d}\log(M/M_{\mathrm{host}})$ and $V_{\mathrm{max}}(M)$ statistics. This is an important selection criterion as merger trees simulated by \textsc{Galacticus} only contain DM halos which are, or previously have been, within the host virial radius, whereas halos outside the virial radius of the host in the cosmological N-Body simulations can also be field halos (i.e. have never passed within the virial radius of the host halo). In other words, it is not proper to directly compare \textsc{Galacticus} simulation results with N-Body without the radius selection criterion. Since the Caterpillar simulation does not include host halos which experienced major mergers (1:3 infall mass ratio) below redshift $z<0.05$, we also exclude halos of this type in \textsc{Galacticus} simulations for our Caterpillar-matched simulations. This is a weak selection criterion and does not have any significant influence on the model calibration results.\par

\begin{figure}
    \centering
    \includegraphics[width=0.45\textwidth]{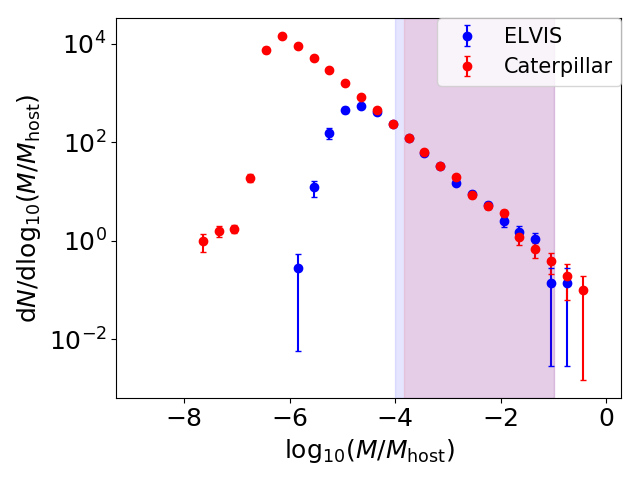}
    \includegraphics[width=0.45\textwidth]{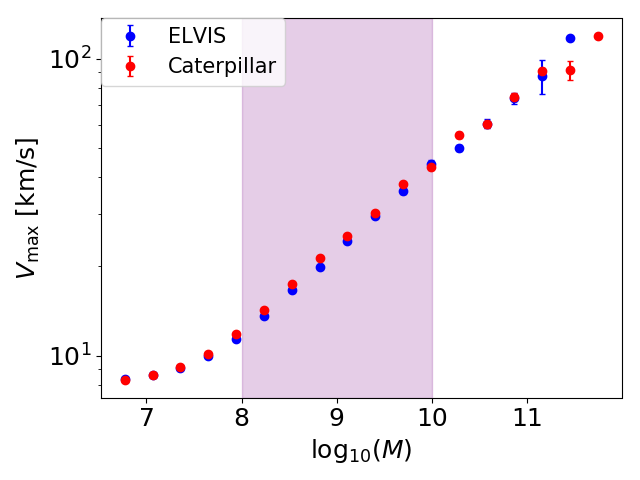}
    \caption{Statistical features of the ELVIS and Caterpillar N-body simulations at $z=0$ used in the model calibration. The top panel shows the sub-halo mass functions and the bottom panel shows $V_{\mathrm{max}}(M)$. Red and blue bands are the mass regions accounted in the MCMC fitting processes for ELVIS and Caterpillar respectively. For the $V_{\mathrm{max}}(M)$ statistics, we exclude all sub-halos with $V_{\mathrm{max}}<8$ km/s.}\label{fig:3}
\end{figure}

Figure \ref{fig:3} shows the sub-halo mass function, $\mathrm{d}N/\mathrm{d}\log(M/M_{\mathrm{host}})$, and maximum circular velocity function, $V_{\mathrm{max}}(M)$, averaged over the 34 (24) host halos in Caterpillar (ELVIS isolated) at $z=0$ respectively. The dots show the mean $\mathrm{d}N/\mathrm{d}\log(M/M_{\mathrm{host}})$ and $V_{\mathrm{max}}(M)$ among all catalogs. Error bars show the error on the mean $\sigma_{\bar{\mathrm{d}}}=\sigma_{d}/\sqrt{N}$,  where $\sigma_\mathrm{d}$ is the standard deviation of N-body data over all host halos, and $N$ is the number of host halos. The host halo mass ranges for ELVIS and Caterpillar simulation are $10^{12}\mathrm{M}_\odot\leq M_{\mathrm{host}}\leq 3\times10^{12}\mathrm{M}_\odot$ and $7\times10^{11}\mathrm{M}_\odot\leq M_{\mathrm{host}}\leq 3\times10^{12}\mathrm{M}_\odot$ respectively, we therefore set identical host mass ranges for \textsc{Galacticus} when generating merger trees. The halo mass resolution of the ELVIS simulation is $2\times10^{7}\mathrm{M}_\odot$, while Caterpillar has a much higher resolution of $6\times10^5\mathrm{M}_\odot$ \footnote{In the ELVIS simulation, a halo is considered to be resolved when it contains more than $100$ particles. In the Caterpillar simulation, an improved halo finder is used and a halo containing more than $20$ particles is considered to be resolved. Applying the same criteria used in ELVIS to Caterpillar, the halo mass resolution of the Caterpillar simulation is $3\times10^6 M_{\odot}$.}. We find that for Caterpillar extending the mass resolution of \textsc{Galacticus} down to $5\times10^{6}\mathrm{M}_\odot$ does not result in significantly stronger constraints on the parameters of our model, but does makes the semi-analytic merger tree construction more computationally expensive. We therefore set the mass resolution of \textsc{Galacticus} to be $M_{\mathrm{res}}=5\times 10^7\mathrm{M}_\odot$ for both ELVIS and Caterpillar fits. We calibrate the non-linear models to $\mathrm{d}N/\mathrm{d}\log(M/M_{\mathrm{host}})$ over fractional mass range $\log_{10}(2M_{\mathrm{res}}/M_{\mathrm{host}}^{\mathrm{min}})\leq \log_{10}(M/M_{\mathrm{host}})<-1$, where $M_{\mathrm{host}}^{\mathrm{min}}$ is the lower limit of the host halo mass distribution. We calibrate models by $V_{\mathrm{max}}(M)$ in sub-halo mass range $\log_{10}(2M_{\mathrm{res}}/\mathrm{M}_\odot)\leq\log_{10}M/\mathrm{M}_\odot<10$ because sub-halos with mass above $10^{10} \mathrm{M}_\odot$ are rare in both ELVIS and Caterpillar simulations, and the $V_{\mathrm{max}}(M)$ statistics for massive satellites are less reliable. ELVIS is complete for sub-halos with $V_{\mathrm{max}}\geq 8$ km/s, while Caterpillar is complete to about $V_{\mathrm{max}}\geq 4$ km/s. To ensure that $V_{\mathrm{max}}(M)$ is not biased by the incompleteness at low masses, we exclude all sub-halos with $V_{\mathrm{max}}<8$ km/s in both \textsc{Galacticus} and N-body simulations when computing the maximum circular velocity function. The blue (red) shaded regions in Figure \ref{fig:3} show the mass ranges we fit for ELVIS (Caterpillar).\par
In order to ensure the statistical errors from the \textsc{Galacticus} simulation are small compared to those contributed by the N-body simulations, we set \textsc{Galacticus} to generate 381 (505) merger trees for ELVIS (Caterpillar), which is about 16 times larger than the corresponding number of N-body simulation merger trees. We therefore ignore the statistical uncertainty contributed by \textsc{Galacticus} simulations when constructing the likelihood function introduced in the following section.\par

\section{MCMC fitting strategy and results}\label{sec:4}
To perform a full search in the $[\alpha$, $\epsilon_h$, $\ln\Lambda]$ 3D parameter space, ideally we would want the MCMC chains to call \textsc{Galacticus} to compute $\mathrm{d}N/\mathrm{d}\log(M/M_{\mathrm{host}})$ and $V_{\mathrm{max}}(M)$ for each new proposed state in the parameter space. However, in this work we use \textsc{Galacticus} to generate 381 (505) merger trees with mass resolution $M_{\mathrm{res}}=5\times10^7\mathrm{M}_\odot$ for ELVIS (Caterpillar) in each simulation, and it takes about 10 CPU hours to evolve the satellites according to the nonlinear evolution models in each simulation. It is not practical to conduct a standard MCMC fitting process
in which each walker may take thousands of steps before convergence is reached. We therefore take an alternative approach. We first select multiple grid points in the 3D parameter space  $[\alpha_i,\epsilon_j,\ln\Lambda_k]$, here $i$, $j$ and $k$ are indexes which run from 1 to $N_x$, with $N_x$ chosen for each parameter $x$, giving a total of $N_\alpha N_\epsilon N_{\ln\Lambda}$ grid points in the parameter space. We then use \textsc{Galacticus} to compute $\mathrm{d}N/\mathrm{d}\log(M/M_{\mathrm{host}})$ as well as $V_{\mathrm{max}}(M)$ for each grid point. \textsc{Galacticus} simulation results for $[\alpha,\epsilon,\ln\Lambda]$ located between grid points are then estimated through linear interpolation. Since $\mathrm{d}N/\mathrm{d}\log(M/M_{\mathrm{host}})$ and $V_{\mathrm{max}}(M)$ change continuously and smoothly under $[\alpha,\epsilon,\ln\Lambda]$ variation, in the limit that the parameter space is gridded infinitely finely the linearly interpolated statistics will be identical to the semi-analytic simulation results.\par
\begin{figure}
    \centering
    \includegraphics[width=0.45\textwidth]{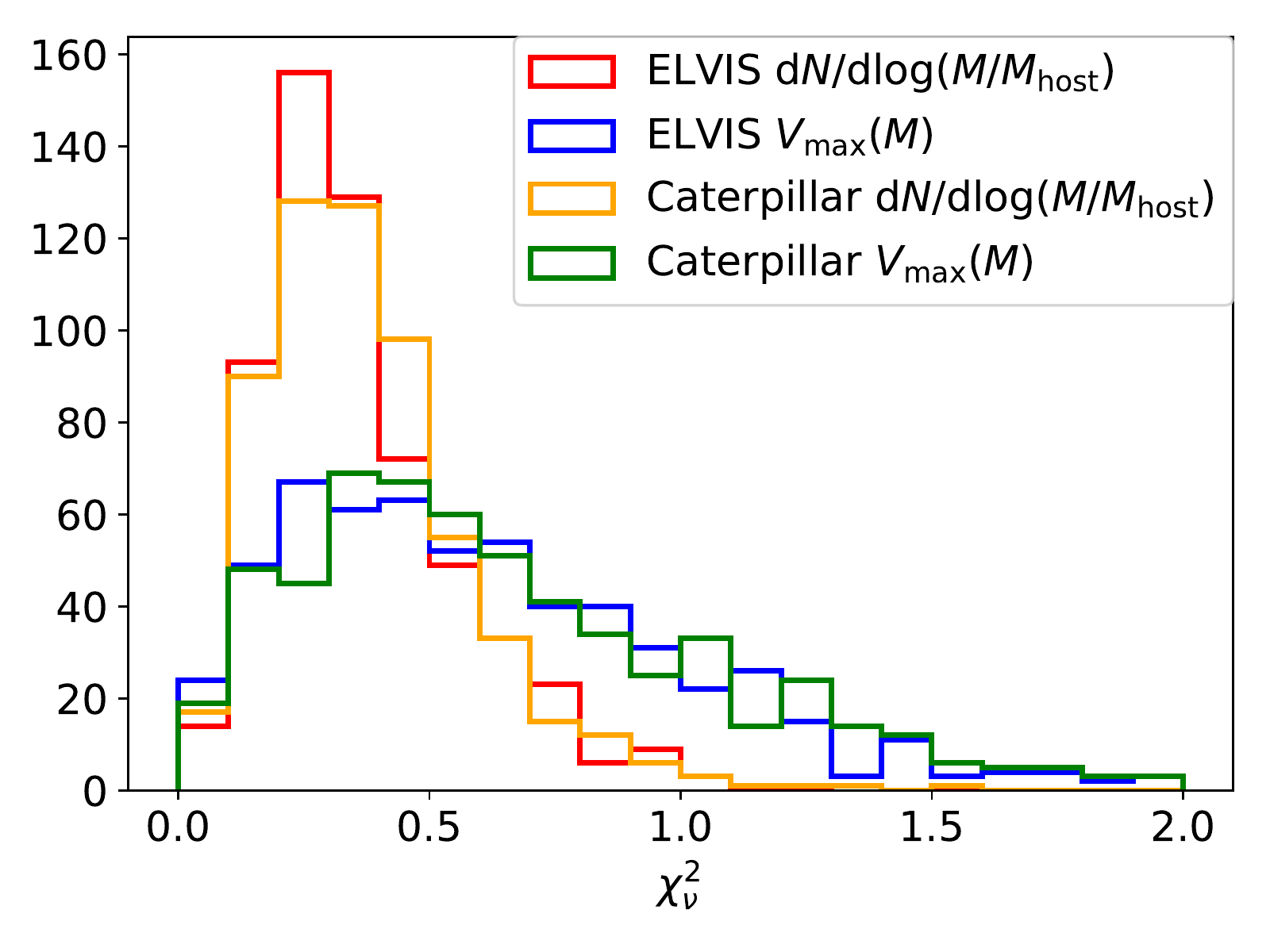}
    \caption{Reduced $\chi^2$ (denoted as $\chi^2_\nu$) distributions of all the tested parameter grid points for the $\gamma=0$ model. For all the cosmologies and statistics we study in this work more than 99\% of the $\chi^2_\nu$ are smaller than 2, meaning that the grid points we take in the parameter space are distributed sufficiently finely that the linearly interpolated $\mathrm{d}N/\mathrm{d}\log(M/M_{\mathrm{host}})$ and $V_{\mathrm{max}}(M)$ agree with \textsc{Galacticus} simulations within the error.}\label{fig:4}
\end{figure}

 
We conduct multiple reduced $\chi^2$ tests to ensure that our gridding of the parameter space is sufficiently fine to give accurate results. Specifically, in each set of tests we remove one grid point of a certain free parameter besides the two grids on the boundaries. For example, if one grid point in the dynamical friction parameter $\alpha$ is removed, $N_{\epsilon}N_{\ln\Lambda}$ grid points will be removed and $(N_\alpha-1)N_{\epsilon}N_{\ln\Lambda}$ grid points will remain in the parameter space. We then linearly interpolate $\mathrm{d}N/\mathrm{d}\log(M/M_{\mathrm{host}})$ and $V_{\mathrm{max}}(M)$ for the removed $N_{\epsilon}N_{\ln\Lambda}$ grid points based on the sub-halo mass functions and maximum velocity functions simulated by \textsc{Galacticus} for the remaining $(N_\alpha-1)N_{\epsilon}N_{\ln\Lambda}$ grid points. Next, we compare the interpolated $\mathrm{d}N/\mathrm{d}\log(M/M_{\mathrm{host}})$ and $V_{\mathrm{max}}(M)$ with those directly simulated by \textsc{Galacticus} for the $N_{\epsilon}N_{\ln\Lambda}$ sets of parameter combinations by computing the reduced $\chi^2$ values:
\begin{equation}
\begin{split}
    \chi^2_\nu&=\left(\sum_i\dfrac{(D_i-D'_i)^2}{\sigma^2_i}\right)/n\,,\\
    \sigma_i^2&=(\sigma_{D}^2)_i+(\sigma_{D'}^2)_i\,,
\end{split}
\end{equation}
here $\chi^2_\nu$ is the reduced $\chi^2$ value, $D$ is the $\mathrm{d}N/\mathrm{d}\log(M/M_{\mathrm{host}})$ or $V_{\mathrm{max}}(M)$ for the removed $N_{\epsilon}N_{\ln\Lambda}$ set of parameter combinations directly simulated by \textsc{Galacticus}, $D'$ is the corresponding $\mathrm{d}N/\mathrm{d}\log(M/M_{\mathrm{host}})$ or $V_{\mathrm{max}}(M)$ linearly interpolated based on statistics of the remaining $(N_\alpha-1)N_{\epsilon}N_{\ln\Lambda}$ grid points, $\sigma_D$ is the error of the mean directly simulated by \textsc{Galacticus}, $\sigma_{D'}$ is estimated through linear interpolation, $i$ is the $M/M_{\mathrm{host}}$ or sub-halo mass bin index, $n$ is the number of bins used in the model calibration. We repeat the above tests for all parameter grid values except those on the boundaries. We find that $99\%$ of the $\chi^2_\nu$ are below 2. Distributions of the $\chi^2_\nu$ for different statistics and cosmologies are presented in Figure \ref{fig:4}. The $\chi^2_\nu$ distribution for the $V_{\mathrm{max}}(M)$ are generally larger than that for the sub-halo mass function because we fit $V_{\mathrm{max}}(M)$ within a smaller mass range, where the \textsc{Galacticus} model uncertainties are larger compared to the $V_{\mathrm{max}}$ predictions for more massive satellites. Combining the model uncertainty with statistical uncertainty, the total errors of $\mathrm{d}N/\mathrm{d}\log(M/M_{\mathrm{host}})$ and $V_{\mathrm{max}}$ predicted by \textsc{Galacticus} are still generally less than half of the N-Body uncertainties. We therefore confirm that our interpolator is a good description of the full model.\par
According to \cite{2016ApJ...830...59L} and \citet{2010MNRAS.406..896B}, the distribution of sub-halo mass functions as well as $V_{\mathrm{max}}(M)$ is non-Gaussian. However, since we compute the average $\mathrm{d}N/\mathrm{d}\log(M/M_{\mathrm{host}})$ and $V_{\mathrm{max}}(M)$ over all host halos in each N-body simulation suite, with 24 (34) host halos in ELVIS (Caterpillar), the central limit theorem suggests that a normal distribution for the mean will be approximately valid. \par

The priors we use in this work are uniform over the range of our gridded parameter space. To locate the prior ranges for the three parameters, we use \textsc{Galacticus} to compute $\mathrm{d}N/\mathrm{d}\log(M/M_{\mathrm{host}})$ or $V_{\mathrm{max}}(M)$ for several points widely distributed throughout the parameter space. Through comparing \textsc{Galacticus} predictions with N-body data we can then roughly determine ranges of individual parameters that produce $\mathrm{d}N/\mathrm{d}\log(M/M_{\mathrm{host}})$ or $V_{\mathrm{max}}(M)$ comparable to N-body statistics. We then take finer grids within the prior ranges and repeat the former process until the prior ranges are narrow but fully cover the potential posteriors of the three parameters. A summary of the prior ranges we use in this work is presented in Table \ref{tb:1}.\par

\begin{table}
\centering
\begin{tabular}{lcc}
\hline
             & $\gamma=0$  &  $\gamma=2.5$  \\
\hline
&&\\[-1em]
$\alpha$     & $(1.5,4.0)$ &  $(2.0,5.0)$   \\
&&\\[-1em]
$\epsilon_h$ & $(0.1,1.5)$ &  $(1.0,12.0)$   \\
&&\\[-1em]
$\ln\Lambda$ & $(0.0,5.0)$ &  $(0.0,8.0)$   \\
&&\\[-1em]
\hline
\end{tabular}
\caption{Summary of uniform prior bounds used in different satellite non-linear evolution models.}\label{tb:1}
\end{table}

Ignoring the adiabatic correction factor in the tidal heating model, for $\gamma=0$ we use a likelihood function:
\begin{equation}\label{eq:17}
\begin{split}
    \ln\mathcal{L}_1&(x|\sigma_x,\alpha,\epsilon_h,\ln\Lambda)=\\
    &-\dfrac{1}{2}\sum\limits_b\left[\dfrac{(x_b-x'_b(\alpha,\epsilon_h,\ln\Lambda))^2}{(\sigma_x^2)_b}+\ln(2\pi (\sigma_x^2)_b)\right]\,,\\
    \ln\mathcal{L}_2&(y|\sigma_y,\alpha,\epsilon_h,\ln\Lambda)=\\
    &-\dfrac{1}{2}\sum\limits_d\left[\dfrac{(y_d-y'_d(\alpha,\epsilon_h,\ln\Lambda))^2}{(\sigma_y^2)_d}+\ln(2\pi (\sigma_y^2)_d)\right]\,,\\
    \ln\mathcal{L}&=\ln\mathcal{L}_1+\ln\mathcal{L}_2.
    \end{split}
\end{equation}
Here $\ln\mathcal{L}_1$ ($\ln\mathcal{L}_2$) is the likelihood function that constrains the sub-halo non-linear evolution models only through the sub-halo mass function (maximum velocity function) statistics. $\ln\mathcal{L}$ is the total likelihood function used for a joint $\mathrm{d}N/\mathrm{d}\log(M/M_{\mathrm{host}})$ and $V_{\mathrm{max}}(M)$ fit. $x$ and $y$ are $\mathrm{d}N/\mathrm{d}\log(M/M_{\mathrm{host}})$ and $V_{\mathrm{max}}(M)$ given by N-body simulation. $x'$ and $y'$ are the interpolated $\mathrm{d}N/\mathrm{d}\log(M/M_{\mathrm{host}})$ and $V_{\mathrm{max}}(M)$ given by \textsc{Galacticus} semi-analytic simulation. $\sigma_x$($\sigma_y$) is the error of the mean of the sub-halo mass function (maximum velocity function) given by N-body simulation. $b$ and $d$ are the index of the fractional mass and sub-halo mass bin located in the MCMC fitting mass ranges that we discussed in Section \ref{sec:3}.\par
For the $\gamma=2.5$ tidal heating model, we find the MCMC fit reduced $\chi^2$ value under the likelihood function of equation~(\ref{eq:17}) is much larger than 1, indicating a severe underestimation of the errors, or that the $\gamma=2.5$ model is not a good description for the N-body data. To study how much the error bars of $\mathrm{d} N/\mathrm{d} \log(M/M_{\mathrm{host}})$ and $V_{\mathrm{max}}(Z)$ should be enlarged to provide a good fit, we replace $\sigma_x$ and $\sigma_y$ in Eq (\ref{eq:17}) by $s_x$ and $s_y$, defined as:
\begin{equation}
    \begin{split}
        (s_x^2)_b&=(\sigma_x^2)_b+f_1^2x_b'(\alpha,\epsilon_h,\ln\Lambda)^2\,,\\
        (s_y^2)_d&=(\sigma_y^2)_d+f_2^2y'_d(\alpha,\epsilon_h,\ln\Lambda)^2\,,
    \end{split}
\end{equation}
here we introduce two additional free parameters $f_1$ and $f_2$ to probe the error bar underestimation for $\mathrm{d} N/\mathrm{d}\log(M/M_{\mathrm{host}})$ and $V_{\mathrm{max}}(M)$ respectively.\par
We use \textsc{emcee} \citep{2013PASP..125..306F} to conduct the MCMC sampling. We run 10 MCMC walkers with initial positions randomly distributed in the gridded parameter space. \par
\begin{figure}
    \centering
    \includegraphics[width=0.45\textwidth]{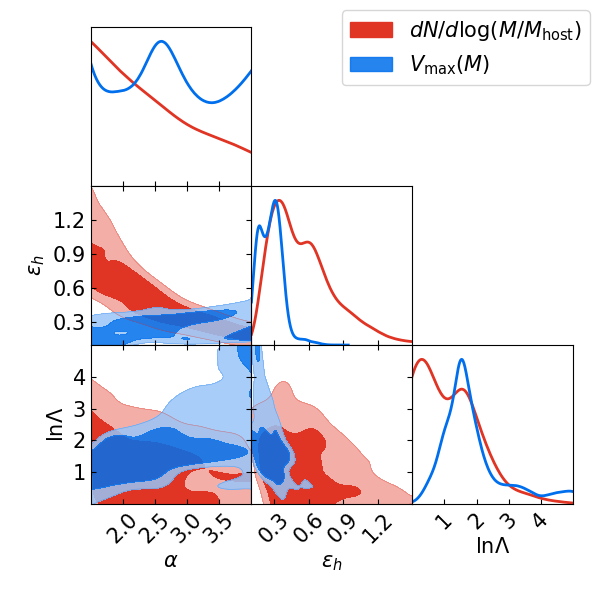}
    \caption{$\alpha,\epsilon_h,\ln\Lambda$ posteriors under adiabatic index $\gamma=0$ and Caterpillar cosmology from the MCMC. The parameters are constrained by either the sub-halo mass function (red) or maximum circular velocity function (blue). $\alpha$ and $\epsilon$ are negatively correlated in the sub-halo mass function, while positively correlated in the maximum circular velocity function. A joint fit for $\mathrm{d}N/\mathrm{d}\log(M/M_{\mathrm{host}})$ and $V_{\mathrm{max}}(M)$ can help to break the tidal stripping and tidal heating model degeneracy, as shown in Figure \ref{fig:6}.}\label{fig:5}
\end{figure}
\begin{figure*}
    \begin{center}
    \begin{tabular}{cc}
    \Large{$\gamma=0$}&\Large{$\gamma=2.5$}\\
    \raisebox{-0.5\height}{\includegraphics[width=0.39\textwidth]{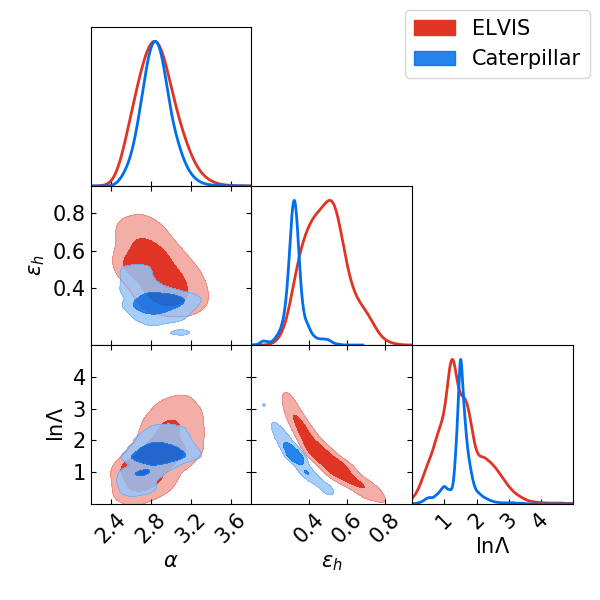}}&\raisebox{-0.5\height}{\includegraphics[width=0.6\textwidth]{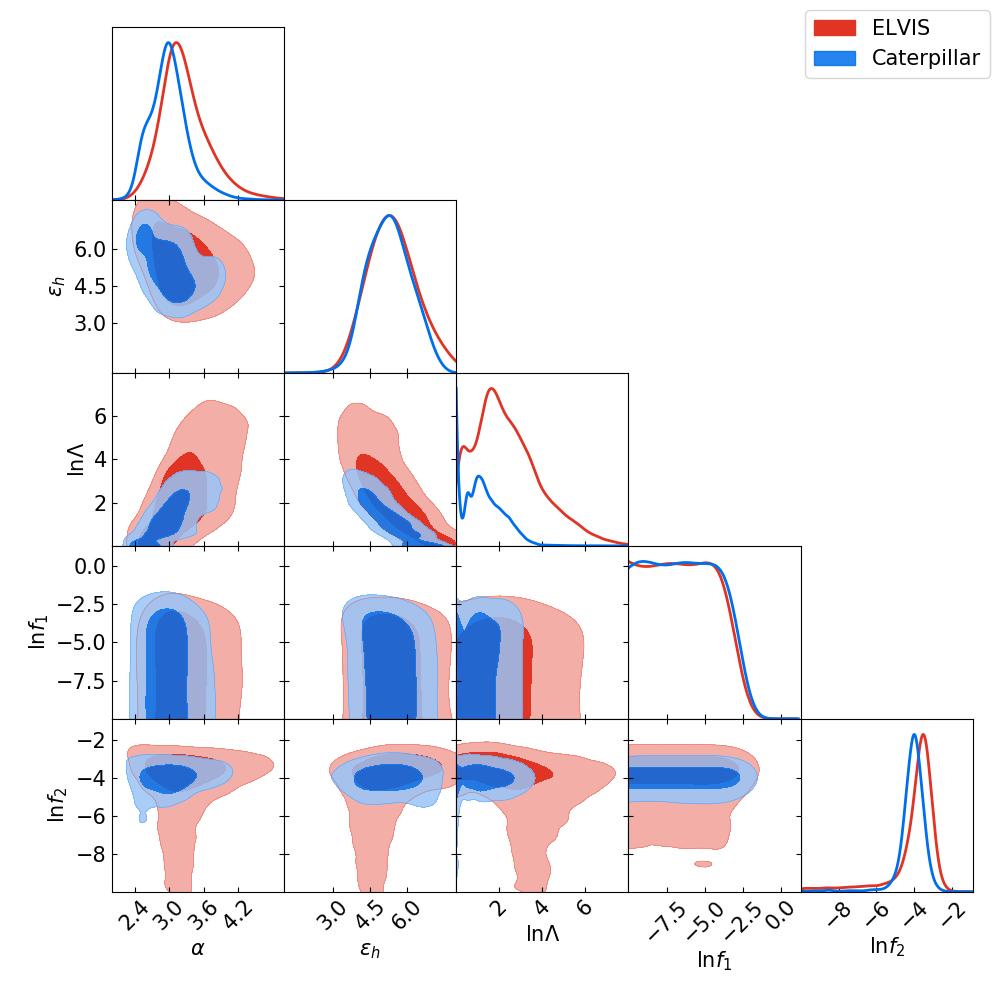}}\\
    \end{tabular}
    \end{center}
    \caption{$\alpha,\epsilon_h,\ln\Lambda$ posteriors under adiabatic index $\gamma=0$ (left panel) and $\gamma=2.5$ (right panel) from MCMC.}\label{fig:6}
\end{figure*}

As an example to show the advantages of combining satellite mass and maximum circular velocity statistics together, we first present the MCMC fitting results using the Caterpillar cosmology and $\gamma=0$ model constrained by $\mathrm{d}N/\mathrm{d}\log(M/M_{\mathrm{host}})$ or $V_{\mathrm{max}}(M)$ alone in Figure~\ref{fig:5}. As  discussed in section \ref{sec:2.4}, $\alpha$ and $\epsilon_h$ are negatively correlated in $\mathrm{d}N/\mathrm{d}\log(M/M_{\mathrm{host}})$ while positively correlated in $V_{\mathrm{max}}(M)$. The $[\alpha,\epsilon_h,\ln\Lambda]$ posteriors of $\gamma=0$ and $\gamma=2.5$ jointly fitted by $\mathrm{d}N/\mathrm{d}\log(M/M_{\mathrm{host}})$ and $V_{\mathrm{max}}(M)$ are shown in Figure~\ref{fig:6}. Comparing with Figure~\ref{fig:5}, the degeneracy between $\alpha$ and $\epsilon_h$ is effectively weakened, and $\ln\Lambda$ is better constrained. The best-fit $[\alpha,\epsilon_h,\ln\Lambda]$ under ELVIS and Caterpillar cosmologies are consistent with each other. The detailed best-fit parameter values and reduced $\chi^2$ test results of Figure~\ref{fig:6} are summarized in Table~\ref{tb:2}. We find that setting the adiabatic correction factor $\gamma$ as 0 and 2.5 gives very different values of the tidal heating coefficient $\epsilon_h$. This is because a positive $\gamma$ weakens the tidal heating rate and a larger $\epsilon_h$ is required to compensate the tidal heating amplitude.\par

\begin{table}
\centering
\begin{tabular}{lcccc}
\hline
              & \multicolumn{2}{c}{$\gamma=0$}                  & \multicolumn{2}{c}{$\gamma=2.5$}                      \\
              & ELVIS                  & Caterpillar            & ELVIS                     & Caterpillar               \\
&&&& \\[-1em]
\hline
&&&& \\[-1em]
$\alpha$      & $2.86_{-0.37}^{+0.39}$ & $2.86_{-0.29}^{+0.33}$ & $3.27_{-0.78}^{+0.89}$     & $3.00_{-0.64}^{+0.68}$   \\
&&&& \\[-1em]
$\epsilon_h$  & $0.49_{-0.21}^{+0.23}$ & $0.33_{-0.11}^{+0.15}$ & $5.4_{-2.0}^{+2.1}$     & $5.3_{-1.6}^{+1.8}$      \\
&&&& \\[-1em]
$\ln\Lambda$  & $1.5_{-1.3}^{+1.5}$    & $1.53_{-0.97}^{+0.93}$ & $2.4_{\downarrow}^{+3.0}$        & $1.3_{\downarrow}^{+1.6}$      \\
&&&& \\[-1em]
$\ln f_1$&-&-&$-6.4_{\downarrow}^{3.4}$&$-6.3_{\downarrow}^{+3.5}$
\\
&&&& \\[-1em]
$\ln f_2$&-&-&$-4.0_{-3.3}^{+1.7}$&$-4.1_{-1.1}^{+1.2}$
\\
&&&& \\[-1em]
$\chi_\nu^2$    & $1.5$                 & $1.0$                 & $1.8$                     & $1.5$                   \\
&&&& \\[-1em]
\hline
\end{tabular}
\caption{Summary of best-fit parameter values and reduced $\chi^2$ of MCMC results shown in Figure \ref{fig:6}. The upper and lower limit for the best-fit parameter values shows the 95\% c.l. $\downarrow$ means the lower limit of the 95\% c.l. reaches the lower bound of the prior.}\label{tb:2}
\end{table}

\begin{figure*}
    \centering
    \includegraphics[width=1\textwidth]{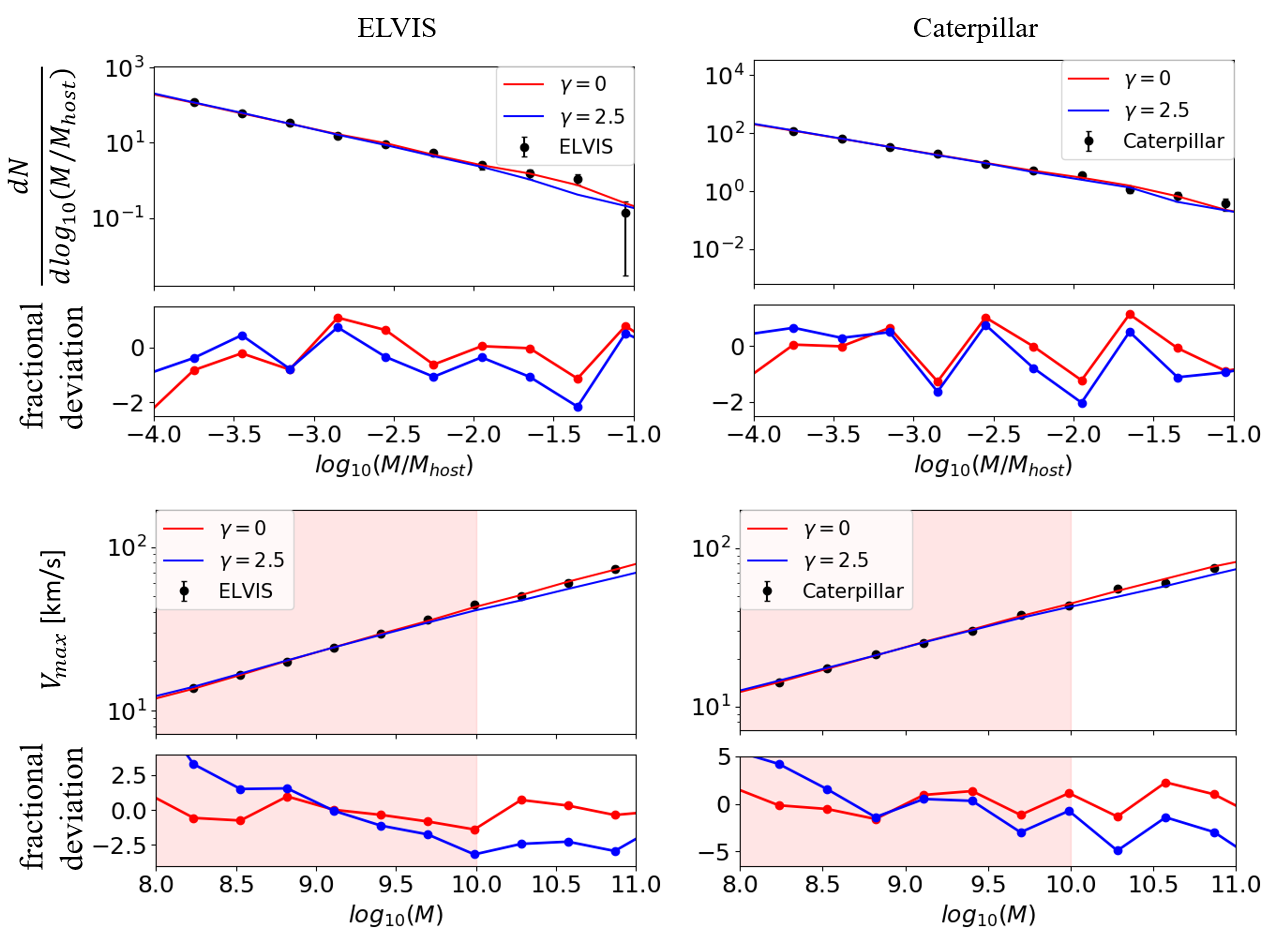}
    \caption{Comparison between  N-body sub-halo statistics and linearly interpolated \textsc{Galacticus} results under best-fit parameters at redshift $z=0$. Figures in the first row present the sub-halo mass function, while those in the second row present the maximum circular velocity function. Figures in the first column are simulated under the ELVIS cosmology while the second column are for the Caterpillar cosmology. The top panel in each figure shows the sub-halo statistics while the bottom panel shows the fractional deviation between the N-Body and best-fit model predictions. Here the fractional deviation is defined as the difference between the statistics predicted by \textsc{Galacticus} and N-body simulations, divided by the error of the mean given by the N-body simulations.}\label{fig:7}
\end{figure*}

\section{discussion}\label{sec:5}

While AP2014 calibrated the non-linear evolution models for sub-halo orbital evolution using only the sub-halo mass function, we add $V_{\mathrm{max}}(M)$ as a further constraint on the free parameters describing  dynamical friction and tidal effect models. The advantage of jointly fitting for $\mathrm{d}N/\mathrm{d}\log(M/M_{\mathrm{host}})$ and $V_{\mathrm{max}}(M)$ is being able to break the degeneracy between $\alpha$ and $\epsilon_h$. \par 
We present the comparison between the sub-halo statistics predicted by \textsc{Galacticus}, evaluated at the best-fit parameters, and the N-body data in Figure~\ref{fig:7}. The  first (second) column shows best-fit \textsc{Galacticus} predictions under the ELVIS (Caterpillar) cosmology. The first (second) row shows the sub-halo mass function (satellite maximum circular velocity) predicted by \textsc{Galacticus} under the calibrated sub-halo evolution models. We present the sub-halo mass function in the mass range used for the MCMC fitting, while the $V_{\mathrm{max}}(M)$ data used for MCMC is enclosed by the red band. In all the subplots of Figure \ref{fig:7}, the top panel presents the sub-halo statistics while the bottom panel shows the fractional error, which is defined as the difference between the statistics predicted by \textsc{Galacticus} and N-Body simulations divided by the error of the mean given by the N-Body simulations. The MCMC $\chi^2_\nu$ tests and the fractional error between N-Body and \textsc{Galacticus} best-fit sub-halo statistics show that the $\gamma=2.5$ tidal heating model fails to reproduce the N-Body $V_{\mathrm{max}}(M)$. We therefore confirm that ignoring the adiabatic correction factor in the tidal heating model, i.e. setting $\gamma=0$, better describes the tidal heating process in CDM N-body simulations.\par 

We identify two possible explanations for the preference of ignoring the adiabatic correction factor in the tidal heating model shown by N-Body simulations. First, since it may take several orbital periods, $T_{\mathrm{orb}}$, before a satellite merges to its host, the position of the satellite DM particle member could undergo a non-negligible change after multiple tidal shocks, breaking the impulse approximation. To account for the break down of the impulse approximation on time scales larger than $T_{\mathrm{orb}}$ we introduce a decaying term $-G_{ab}(t')T_{\mathrm{orb}}$ in the time integral of equation~(\ref{eq:tidal_int}). The decaying term effectively suppresses the tidal heating rate and plays a similar role as the adiabatic correction factor.  Therefore the presence of the decaying factor might be the cause of a trivial adiabatic correction factor, i.e., $\gamma=0$. We leave a more careful comparison between the decaying term of the tidal tensor time integral and the adiabatic correction factor to future work. As a second possible explanation, \cite{2018MNRAS.474.3043V} show that in the cosmological Bolshoi simulation, the overall impact of the adiabatic correction factor on the energy injected to sub-halos by tidal heating effect is negligible. Moreover, for sub-halos with orbital circularity $\eta\gtrsim0.2$, the impulse approximation combined with the adiabatic correction factor underestimates the sub-halo mass fraction stripped off by the tidal effects. Therefore, setting $\gamma=0$ effectively enhances tidal heating and helps to compensate the underestimation of tidal effects.\par

\section{Conclusion}\label{sec:6}
In this work we develop a fast MCMC fitting strategy for \textsc{Galacticus} sub-halo orbital evolution models. We apply this new MCMC method to fit three parameters related to the dynamical friction, tidal stripping, and tidal heating models introduced to \textsc{Galacticus} by AP2014. We show that sub-halo statistics predicted by \textsc{Galacticus} are in good agreement with ELVIS and Caterpillar N-body simulations. \par
Since both tidal stripping and tidal heating effects increase the mass loss from satellites, we find that using the sub-halo mass function alone for model calibration leads to a degeneracy between tidal effects. We show that including $V_{\mathrm{max}}(M)$, which is sensitive to the sub-halo density profile, can break this degeneracy. \par 
Limited by a lack of massive substructures in ELVIS and Caterpillar N-body simulations, we fail to place a strong constraint on the dynamical friction model, which mostly influences massive sub-halos. Other simulations and statistics might be helpful to break the negative degeneracy between dynamical friction and tidal heating effects. First, future N-Body simulations or current cluster zoom-in simulations (e.g. \cite{2013MNRAS.429..323S,2018MNRAS.480.2898C}) with large halo sample volumes will contain a larger number count of massive sub-halos and provide tighter constraint on the dynamical friction model. Second, dynamical friction can be probed in more detail through placing a massive sub-halo in the host halo and tracking its orbital evolution. Moreover, strong dynamical friction increases the concentration of sub-halos toward the host halo center. Therefore, the radial distribution of sub-halos may help to place stronger constraints on the dynamical friction model.  We plan to explore these possibilities in the future.\par 

We find evidence from the maximum circular velocities, $V_{\mathrm{max}}(M)$, predicted by \textsc{Galacticus} with the calibrated sub-halo evolution models that ignoring the adiabatic correction factor in the tidal heating model better describes the cosmological simulation data than the original $\gamma=2.5$ model of \cite{1999ApJ...513..626G}. It is possible that the decaying term we introduce to the time integral of tidal tensor in the tidal heating model effectively acts to replace some of the adiabatic correction factor. Alternatively, tidal heating with non-zero adiabatic correction may only be a good description for sub-halos with more radial orbits and may therefore underestimate the averaged tidal heating effects throughout the sub-halo population. Extracting the tidal heating energy directly from N-body simulation will be helpful to break the degeneracy between the tidal tensor decaying term and adiabatic correction factor. For $\gamma=0$, MCMC gives the best-fit strength of dynamical friction, tidal stripping, and tidal heating effects as $\ln\Lambda=1.5^{+1.5}_{-1.3}$, $\alpha=2.86^{+0.39}_{-0.37}$, $\epsilon_h=0.49^{+0.23}_{-0.21}$ for ELVIS cosmology and $\ln\Lambda=1.53^{+0.93}_{-0.97}$, $\alpha=2.86^{+0.33}_{-0.29}$, $\epsilon_h=0.33^{+0.15}_{-0.11}$ for Caterpillar at 95\% c.l. These posteriors agree within the 95\% c.l.\par
In this work we only calibrate the non-linear sub-halo evolution models to the isolated N-Body halo catalogs and do not account for the interaction among host halos, our best-fit result is therefore applicable to the dark matter substructure evolution within isolated host halos. Although both ELVIS and Caterpillar simulations are focused on Milky Way-sized halo with mass of about $10^{12}\mathrm{M}_\odot$, our best-fit result is also applicable to dark matter substructure evolution within host halos with different masses as gravity is scale invariant.\par
A good, quantitative understanding of DM substructure evolution is crucial for constraining DM properties with future observations. The best-fit results of this work can make accurate and fast predictions for the sub-halo populations based on physics models and provide priors for future DM substructure studies and measurements. Orbital evolution models for DM sub-halos are still under intensive study and the best fit values of the parameters may vary with additional model refinements. Our fast MCMC fitting framework will be applicable to more sophisticated sub-halo and satellite evolution models in the future.\par 
\section{Acknowledgement}
We wish to thank Shenglong Wang for the IT support. We thank Brendan Griffen for providing Caterpillar simulation catalogs with $\mathrm{LX} = 14$ resolution. S.Y thanks Yiyang Wu for useful discussions about MCMC fitting strategies. X.D. thanks Ethan Nadler for beneficial discussions on sub-halo mass function and maximum velocity function in N-body simulations and SAMs.  This material is based on work supported by NASA under award numbers 80NSSC18K1014 and NNH17ZDA001N.  This work was supported by the Simons Foundation.

\appendix
\section{initial velocity distribution of satellites}\label{apdx:1}
During the \textsc{Galacticus} simulation, the initial velocity of each sub-halo is randomly drawn from a distribution. This velocity distribution therefore determines the subsequent evolution of sub-halo properties within the host. In this work, we apply the most up-to-date velocity distribution given by \cite{2015MNRAS.448.1674J} (here after Jiang2015), which is fitted to the cosmological N-Body simulation ``DOVE'', which has DM halo mass resolution $M_{\mathrm{res}}=2\times 10^8 \mathrm{M}_\odot$. To show how the choice of satellite initial velocity distribution influences our calibration results, in this section we use the velocity distribution provided by \cite{2005MNRAS.358..551B} (here after Benson2005) instead and compare the best-fit non-linear sub-halo evolution model parameters $[\ln\Lambda,\alpha,\epsilon_h]$ with those fitted under the Jiang2015 initial velocity distribution.\par
\begin{figure*}
    \centering
    \includegraphics[width=1.0\columnwidth]{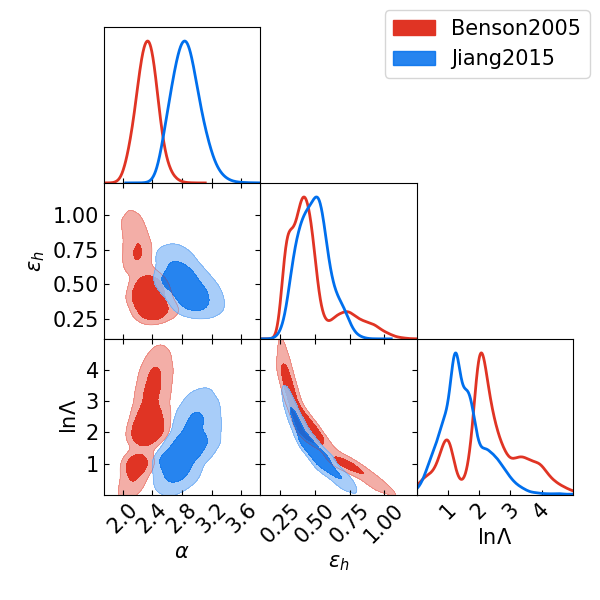}
    \includegraphics[width=1.0\columnwidth]{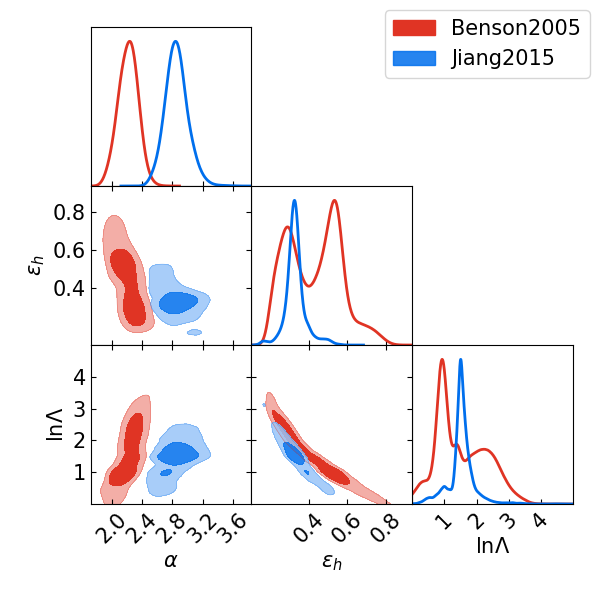}
    \caption{$\alpha$, $\epsilon_h$ and $\ln\Lambda$ posteriors with ELVIS (left) and Caterpillar (right) cosmologies, adiabatic index $\gamma=0$ and with variation of the satellite initial velocity distribution. Red posteriors corresponds to the initial satellite velocity distribution given by Benson2005~\citep{2005MNRAS.358..551B}. The Blue posterior corresponds to the initial velocity distribution given by Jiang2015~\citep{2015MNRAS.448.1674J}}\label{fig:A1}
\end{figure*}

Taking the adiabatic correction factor $\gamma=0$ case as an example, comparisons between the non-linear evolution model best-fit values and posteriors under Benson2005 and Jiang2015 velocity distributions are summarized in Figure~\ref{fig:A1}. For both Caterpillar and ELVIS cosmologies, the best-fit results of $\epsilon_h$ and $\ln\Lambda$ with the Jiang2015 and Benson2005 velocity distributions agree within 95\% confidence level, but the mean of the tidal stripping mass loss rate $\alpha$ shifts more than $2\sigma$. Further tests show that this velocity distribution variation can lead to a significant shift of the sub-halo mass function $\mathrm{d}N/\mathrm{d}\log(M/M_{\mathrm{host}})$ and the maximum velocity function $V_{\mathrm{max}}(M)$ compared to the statistical uncertainty.\par
We believe the velocity distribution provided by Jiang2015 is a better choice for this work because the DOVE N-Body simulation has mass resolution similar to the one we apply to the \textsc{Galacticus} simulations, while the Benson2005 satellite infall velocity distribution is fitted to N-Body simulations with lower mass resolution. Although the velocity distribution variation could bring a $2\sigma$ shift to the best-fit $\alpha$, it does not influence any of our qualitative conclusions.
\par

\section*{Data availability}
The data used to support the findings of this study are available from the corresponding author upon request.




\bibliographystyle{mnras}
\bibliography{Galacticus}

\begin{thebibliography}{}
\makeatletter
\relax
\def\mn@urlcharsother{\let\do\@makeother \do\$\do\&\do\#\do\^\do\_\do\%\do\~}
\def\mn@doi{\begingroup\mn@urlcharsother \@ifnextchar [ {\mn@doi@}
  {\mn@doi@[]}}
\def\mn@doi@[#1]#2{\def\@tempa{#1}\ifx\@tempa\@empty \href
  {http://dx.doi.org/#2} {doi:#2}\else \href {http://dx.doi.org/#2} {#1}\fi
  \endgroup}
\def\mn@eprint#1#2{\mn@eprint@#1:#2::\@nil}
\def\mn@eprint@arXiv#1{\href {http://arxiv.org/abs/#1} {{\tt arXiv:#1}}}
\def\mn@eprint@dblp#1{\href {http://dblp.uni-trier.de/rec/bibtex/#1.xml}
  {dblp:#1}}
\def\mn@eprint@#1:#2:#3:#4\@nil{\def\@tempa {#1}\def\@tempb {#2}\def\@tempc
  {#3}\ifx \@tempc \@empty \let \@tempc \@tempb \let \@tempb \@tempa \fi \ifx
  \@tempb \@empty \def\@tempb {arXiv}\fi \@ifundefined
  {mn@eprint@\@tempb}{\@tempb:\@tempc}{\expandafter \expandafter \csname
  mn@eprint@\@tempb\endcsname \expandafter{\@tempc}}}

\bibitem[\protect\citeauthoryear{{Ahn} \& {Shapiro}}{{Ahn} \&
  {Shapiro}}{2005}]{2005MNRAS.363.1092A}
{Ahn} K.,  {Shapiro} P.~R.,  2005, \mn@doi [\mnras]
  {10.1111/j.1365-2966.2005.09492.x}, \href
  {https://ui.adsabs.harvard.edu/abs/2005MNRAS.363.1092A} {363, 1092}

\bibitem[\protect\citeauthoryear{{Anderson} et~al.,}{{Anderson}
  et~al.}{2012}]{2012MNRAS.427.3435A}
{Anderson} L.,  et~al., 2012, \mn@doi [\mnras]
  {10.1111/j.1365-2966.2012.22066.x}, \href
  {https://ui.adsabs.harvard.edu/abs/2012MNRAS.427.3435A} {427, 3435}

\bibitem[\protect\citeauthoryear{{Banik}, {Bertone}, {Bovy}  \&
  {Bozorgnia}}{{Banik} et~al.}{2018}]{2018JCAP...07..061B}
{Banik} N.,  {Bertone} G.,  {Bovy} J.,   {Bozorgnia} N.,  2018, \mn@doi [\jcap]
  {10.1088/1475-7516/2018/07/061}, \href
  {https://ui.adsabs.harvard.edu/abs/2018JCAP...07..061B} {2018, 061}

\bibitem[\protect\citeauthoryear{{Banik}, {Bovy}, {Bertone}, {Erkal}  \& {de
  Boer}}{{Banik} et~al.}{2019}]{2019arXiv191102663B}
{Banik} N.,  {Bovy} J.,  {Bertone} G.,  {Erkal} D.,   {de Boer} T.~J.~L.,
  2019, arXiv e-prints, \href
  {https://ui.adsabs.harvard.edu/abs/2019arXiv191102663B} {p. arXiv:1911.02663}

\bibitem[\protect\citeauthoryear{{Benson}}{{Benson}}{2005}]{2005MNRAS.358..551B}
{Benson} A.~J.,  2005, \mn@doi [\mnras] {10.1111/j.1365-2966.2005.08788.x},
  \href {https://ui.adsabs.harvard.edu/abs/2005MNRAS.358..551B} {358, 551}

\bibitem[\protect\citeauthoryear{{Benson}}{{Benson}}{2012}]{2012NewA...17..175B}
{Benson} A.~J.,  2012, \mn@doi [\na] {10.1016/j.newast.2011.07.004}, \href
  {https://ui.adsabs.harvard.edu/abs/2012NewA...17..175B} {17, 175}

\bibitem[\protect\citeauthoryear{{Benson}, {Lacey}, {Baugh}, {Cole}  \&
  {Frenk}}{{Benson} et~al.}{2002}]{2002MNRAS.333..156B}
{Benson} A.~J.,  {Lacey} C.~G.,  {Baugh} C.~M.,  {Cole} S.,   {Frenk} C.~S.,
  2002, \mn@doi [\mnras] {10.1046/j.1365-8711.2002.05387.x}, \href
  {https://ui.adsabs.harvard.edu/abs/2002MNRAS.333..156B} {333, 156}

\bibitem[\protect\citeauthoryear{{Benson} et~al.,}{{Benson}
  et~al.}{2013}]{2013MNRAS.428.1774B}
{Benson} A.~J.,  et~al., 2013, \mn@doi [\mnras] {10.1093/mnras/sts159}, \href
  {https://ui.adsabs.harvard.edu/abs/2013MNRAS.428.1774B} {428, 1774}

\bibitem[\protect\citeauthoryear{{Birrer}, {Amara}  \& {Refregier}}{{Birrer}
  et~al.}{2017}]{2017JCAP...05..037B}
{Birrer} S.,  {Amara} A.,   {Refregier} A.,  2017, \mn@doi [\jcap]
  {10.1088/1475-7516/2017/05/037}, \href
  {https://ui.adsabs.harvard.edu/abs/2017JCAP...05..037B} {2017, 037}

\bibitem[\protect\citeauthoryear{{Boehm} \& {Schaeffer}}{{Boehm} \&
  {Schaeffer}}{2005}]{2005A&A...438..419B}
{Boehm} C.,  {Schaeffer} R.,  2005, \mn@doi [\aap]
  {10.1051/0004-6361:20042238}, \href
  {https://ui.adsabs.harvard.edu/abs/2005A&A...438..419B} {438, 419}

\bibitem[\protect\citeauthoryear{{Bonaca} \& {Hogg}}{{Bonaca} \&
  {Hogg}}{2018}]{2018ApJ...867..101B}
{Bonaca} A.,  {Hogg} D.~W.,  2018, \mn@doi [\apj] {10.3847/1538-4357/aae4da},
  \href {https://ui.adsabs.harvard.edu/abs/2018ApJ...867..101B} {867, 101}

\bibitem[\protect\citeauthoryear{{Bonaca} et~al.,}{{Bonaca}
  et~al.}{2020}]{2020ApJ...892L..37B}
{Bonaca} A.,  et~al., 2020, \mn@doi [\apjl] {10.3847/2041-8213/ab800c}, \href
  {https://ui.adsabs.harvard.edu/abs/2020ApJ...892L..37B} {892, L37}

\bibitem[\protect\citeauthoryear{{Bond}, {Cole}, {Efstathiou}  \&
  {Kaiser}}{{Bond} et~al.}{1991}]{1991ApJ...379..440B}
{Bond} J.~R.,  {Cole} S.,  {Efstathiou} G.,   {Kaiser} N.,  1991, \mn@doi
  [\apj] {10.1086/170520}, \href
  {https://ui.adsabs.harvard.edu/abs/1991ApJ...379..440B} {379, 440}

\bibitem[\protect\citeauthoryear{{Bosma}}{{Bosma}}{1981}]{1981AJ.....86.1825B}
{Bosma} A.,  1981, \mn@doi [\aj] {10.1086/113063}, \href
  {https://ui.adsabs.harvard.edu/abs/1981AJ.....86.1825B} {86, 1825}

\bibitem[\protect\citeauthoryear{{Bovy}, {Erkal}  \& {Sanders}}{{Bovy}
  et~al.}{2017}]{2017MNRAS.466..628B}
{Bovy} J.,  {Erkal} D.,   {Sanders} J.~L.,  2017, \mn@doi [\mnras]
  {10.1093/mnras/stw3067}, \href
  {https://ui.adsabs.harvard.edu/abs/2017MNRAS.466..628B} {466, 628}

\bibitem[\protect\citeauthoryear{{Bower}}{{Bower}}{1991}]{1991MNRAS.248..332B}
{Bower} R.~G.,  1991, \mn@doi [\mnras] {10.1093/mnras/248.2.332}, \href
  {https://ui.adsabs.harvard.edu/abs/1991MNRAS.248..332B} {248, 332}

\bibitem[\protect\citeauthoryear{{Boylan-Kolchin}, {Springel}, {White}  \&
  {Jenkins}}{{Boylan-Kolchin} et~al.}{2010}]{2010MNRAS.406..896B}
{Boylan-Kolchin} M.,  {Springel} V.,  {White} S. D.~M.,   {Jenkins} A.,  2010,
  \mn@doi [\mnras] {10.1111/j.1365-2966.2010.16774.x}, \href
  {https://ui.adsabs.harvard.edu/abs/2010MNRAS.406..896B} {406, 896}

\bibitem[\protect\citeauthoryear{{Boylan-Kolchin}, {Bullock}  \&
  {Kaplinghat}}{{Boylan-Kolchin} et~al.}{2011}]{2011MNRAS.415L..40B}
{Boylan-Kolchin} M.,  {Bullock} J.~S.,   {Kaplinghat} M.,  2011, \mn@doi
  [\mnras] {10.1111/j.1745-3933.2011.01074.x}, \href
  {https://ui.adsabs.harvard.edu/abs/2011MNRAS.415L..40B} {415, L40}

\bibitem[\protect\citeauthoryear{{Boylan-Kolchin}, {Bullock}  \&
  {Kaplinghat}}{{Boylan-Kolchin} et~al.}{2012}]{2012MNRAS.422.1203B}
{Boylan-Kolchin} M.,  {Bullock} J.~S.,   {Kaplinghat} M.,  2012, \mn@doi
  [\mnras] {10.1111/j.1365-2966.2012.20695.x}, \href
  {https://ui.adsabs.harvard.edu/abs/2012MNRAS.422.1203B} {422, 1203}

\bibitem[\protect\citeauthoryear{{Bozek} et~al.,}{{Bozek}
  et~al.}{2019}]{2019MNRAS.483.4086B}
{Bozek} B.,  et~al., 2019, \mn@doi [\mnras] {10.1093/mnras/sty3300}, \href
  {https://ui.adsabs.harvard.edu/abs/2019MNRAS.483.4086B} {483, 4086}

\bibitem[\protect\citeauthoryear{{Brooks} \& {Zolotov}}{{Brooks} \&
  {Zolotov}}{2014}]{2014ApJ...786...87B}
{Brooks} A.~M.,  {Zolotov} A.,  2014, \mn@doi [\apj]
  {10.1088/0004-637X/786/2/87}, \href
  {https://ui.adsabs.harvard.edu/abs/2014ApJ...786...87B} {786, 87}

\bibitem[\protect\citeauthoryear{{Bullock}}{{Bullock}}{2010}]{2010arXiv1009.4505B}
{Bullock} J.~S.,  2010, arXiv e-prints, \href
  {https://ui.adsabs.harvard.edu/abs/2010arXiv1009.4505B} {p. arXiv:1009.4505}

\bibitem[\protect\citeauthoryear{{Bullock} \& {Boylan-Kolchin}}{{Bullock} \&
  {Boylan-Kolchin}}{2017}]{2017ARA&A..55..343B}
{Bullock} J.~S.,  {Boylan-Kolchin} M.,  2017, \mn@doi [\araa]
  {10.1146/annurev-astro-091916-055313}, \href
  {https://ui.adsabs.harvard.edu/abs/2017ARA&A..55..343B} {55, 343}

\bibitem[\protect\citeauthoryear{{Buschmann}, {Kopp}, {Safdi}  \&
  {Wu}}{{Buschmann} et~al.}{2018}]{2018PhRvL.120u1101B}
{Buschmann} M.,  {Kopp} J.,  {Safdi} B.~R.,   {Wu} C.-L.,  2018, \mn@doi [\prl]
  {10.1103/PhysRevLett.120.211101}, \href
  {https://ui.adsabs.harvard.edu/abs/2018PhRvL.120u1101B} {120, 211101}

\bibitem[\protect\citeauthoryear{{Chandrasekhar}}{{Chandrasekhar}}{1943}]{1943RvMP...15....1C}
{Chandrasekhar} S.,  1943, \mn@doi [Reviews of Modern Physics]
  {10.1103/RevModPhys.15.1}, \href
  {https://ui.adsabs.harvard.edu/abs/1943RvMP...15....1C} {15, 1}

\bibitem[\protect\citeauthoryear{{Cui} et~al.,}{{Cui}
  et~al.}{2018}]{2018MNRAS.480.2898C}
{Cui} W.,  et~al., 2018, \mn@doi [\mnras] {10.1093/mnras/sty2111}, \href
  {https://ui.adsabs.harvard.edu/abs/2018MNRAS.480.2898C} {480, 2898}

\bibitem[\protect\citeauthoryear{{Diemer} \& {Kravtsov}}{{Diemer} \&
  {Kravtsov}}{2015}]{2015ApJ...799..108D}
{Diemer} B.,  {Kravtsov} A.~V.,  2015, \mn@doi [\apj]
  {10.1088/0004-637X/799/1/108}, \href
  {https://ui.adsabs.harvard.edu/abs/2015ApJ...799..108D} {799, 108}

\bibitem[\protect\citeauthoryear{{Donato}, {Gentile}  \& {Salucci}}{{Donato}
  et~al.}{2004}]{2004MNRAS.353L..17D}
{Donato} F.,  {Gentile} G.,   {Salucci} P.,  2004, \mn@doi [\mnras]
  {10.1111/j.1365-2966.2004.08220.x}, \href
  {https://ui.adsabs.harvard.edu/abs/2004MNRAS.353L..17D} {353, L17}

\bibitem[\protect\citeauthoryear{{Donato} et~al.,}{{Donato}
  et~al.}{2009}]{2009MNRAS.397.1169D}
{Donato} F.,  et~al., 2009, \mn@doi [\mnras]
  {10.1111/j.1365-2966.2009.15004.x}, \href
  {https://ui.adsabs.harvard.edu/abs/2009MNRAS.397.1169D} {397, 1169}

\bibitem[\protect\citeauthoryear{{Erkal}, {Belokurov}, {Bovy}  \& {Sand
  ers}}{{Erkal} et~al.}{2016}]{2016MNRAS.463..102E}
{Erkal} D.,  {Belokurov} V.,  {Bovy} J.,   {Sand ers} J.~L.,  2016, \mn@doi
  [\mnras] {10.1093/mnras/stw1957}, \href
  {https://ui.adsabs.harvard.edu/abs/2016MNRAS.463..102E} {463, 102}

\bibitem[\protect\citeauthoryear{{Foreman-Mackey}, {Hogg}, {Lang}  \&
  {Goodman}}{{Foreman-Mackey} et~al.}{2013}]{2013PASP..125..306F}
{Foreman-Mackey} D.,  {Hogg} D.~W.,  {Lang} D.,   {Goodman} J.,  2013, \mn@doi
  [\pasp] {10.1086/670067}, \href
  {https://ui.adsabs.harvard.edu/abs/2013PASP..125..306F} {125, 306}

\bibitem[\protect\citeauthoryear{{Garavito-Camargo}, {Besla}, {Laporte},
  {Johnston}, {G{\'o}mez}  \& {Watkins}}{{Garavito-Camargo}
  et~al.}{2019}]{2019ApJ...884...51G}
{Garavito-Camargo} N.,  {Besla} G.,  {Laporte} C. F.~P.,  {Johnston} K.~V.,
  {G{\'o}mez} F.~A.,   {Watkins} L.~L.,  2019, \mn@doi [\apj]
  {10.3847/1538-4357/ab32eb}, \href
  {https://ui.adsabs.harvard.edu/abs/2019ApJ...884...51G} {884, 51}

\bibitem[\protect\citeauthoryear{{Garrison-Kimmel}, {Boylan-Kolchin}, {Bullock}
   \& {Lee}}{{Garrison-Kimmel} et~al.}{2014}]{2014MNRAS.438.2578G}
{Garrison-Kimmel} S.,  {Boylan-Kolchin} M.,  {Bullock} J.~S.,   {Lee} K.,
  2014, \mn@doi [\mnras] {10.1093/mnras/stt2377}, \href
  {https://ui.adsabs.harvard.edu/abs/2014MNRAS.438.2578G} {438, 2578}

\bibitem[\protect\citeauthoryear{{Gilman}, {Birrer}, {Treu}, {Nierenberg}  \&
  {Benson}}{{Gilman} et~al.}{2019}]{2019MNRAS.487.5721G}
{Gilman} D.,  {Birrer} S.,  {Treu} T.,  {Nierenberg} A.,   {Benson} A.,  2019,
  \mn@doi [\mnras] {10.1093/mnras/stz1593}, \href
  {https://ui.adsabs.harvard.edu/abs/2019MNRAS.487.5721G} {487, 5721}

\bibitem[\protect\citeauthoryear{{Gilman}, {Birrer}, {Nierenberg}, {Treu}, {Du}
   \& {Benson}}{{Gilman} et~al.}{2020}]{2020MNRAS.491.6077G}
{Gilman} D.,  {Birrer} S.,  {Nierenberg} A.,  {Treu} T.,  {Du} X.,   {Benson}
  A.,  2020, \mn@doi [\mnras] {10.1093/mnras/stz3480}, \href
  {https://ui.adsabs.harvard.edu/abs/2020MNRAS.491.6077G} {491, 6077}

\bibitem[\protect\citeauthoryear{{Gnedin} \& {Ostriker}}{{Gnedin} \&
  {Ostriker}}{1999}]{1999ApJ...513..626G}
{Gnedin} O.~Y.,  {Ostriker} J.~P.,  1999, \mn@doi [\apj] {10.1086/306864},
  \href {https://ui.adsabs.harvard.edu/abs/1999ApJ...513..626G} {513, 626}

\bibitem[\protect\citeauthoryear{{Gnedin}, {Hernquist}  \& {Ostriker}}{{Gnedin}
  et~al.}{1997}]{1997astro.ph..9161G}
{Gnedin} O.~Y.,  {Hernquist} L.,   {Ostriker} J.~P.,  1997, arXiv e-prints,
  \href {https://ui.adsabs.harvard.edu/abs/1997astro.ph..9161G} {pp
  astro--ph/9709161}

\bibitem[\protect\citeauthoryear{{Griffen}, {Ji}, {Dooley}, {G{\'o}mez},
  {Vogelsberger}, {O'Shea}  \& {Frebel}}{{Griffen}
  et~al.}{2016}]{2016ApJ...818...10G}
{Griffen} B.~F.,  {Ji} A.~P.,  {Dooley} G.~A.,  {G{\'o}mez} F.~A.,
  {Vogelsberger} M.,  {O'Shea} B.~W.,   {Frebel} A.,  2016, \mn@doi [\apj]
  {10.3847/0004-637X/818/1/10}, \href
  {https://ui.adsabs.harvard.edu/abs/2016ApJ...818...10G} {818, 10}

\bibitem[\protect\citeauthoryear{{Hezaveh}, {Dalal}, {Holder}, {Kisner},
  {Kuhlen}  \& {Perreault Levasseur}}{{Hezaveh}
  et~al.}{2016}]{2016JCAP...11..048H}
{Hezaveh} Y.,  {Dalal} N.,  {Holder} G.,  {Kisner} T.,  {Kuhlen} M.,
  {Perreault Levasseur} L.,  2016, \mn@doi [\jcap]
  {10.1088/1475-7516/2016/11/048}, \href
  {https://ui.adsabs.harvard.edu/abs/2016JCAP...11..048H} {2016, 048}

\bibitem[\protect\citeauthoryear{{Hsueh}, {Enzi}, {Vegetti}, {Auger},
  {Fassnacht}, {Despali}, {Koopmans}  \& {McKean}}{{Hsueh}
  et~al.}{2020}]{2020MNRAS.492.3047H}
{Hsueh} J.~W.,  {Enzi} W.,  {Vegetti} S.,  {Auger} M.~W.,  {Fassnacht} C.~D.,
  {Despali} G.,  {Koopmans} L.~V.~E.,   {McKean} J.~P.,  2020, \mnras, \href
  {https://ui.adsabs.harvard.edu/abs/2020MNRAS.492.3047H} {492, 3047}

\bibitem[\protect\citeauthoryear{{Ibata}, {Malhan}  \& {Martin}}{{Ibata}
  et~al.}{2019}]{2019ApJ...872..152I}
{Ibata} R.~A.,  {Malhan} K.,   {Martin} N.~F.,  2019, \mn@doi [\apj]
  {10.3847/1538-4357/ab0080}, \href
  {https://ui.adsabs.harvard.edu/abs/2019ApJ...872..152I} {872, 152}

\bibitem[\protect\citeauthoryear{{Jiang}, {Cole}, {Sawala}  \& {Frenk}}{{Jiang}
  et~al.}{2015}]{2015MNRAS.448.1674J}
{Jiang} L.,  {Cole} S.,  {Sawala} T.,   {Frenk} C.~S.,  2015, \mn@doi [\mnras]
  {10.1093/mnras/stv053}, \href
  {https://ui.adsabs.harvard.edu/abs/2015MNRAS.448.1674J} {448, 1674}

\bibitem[\protect\citeauthoryear{{Kaplinghat}, {Tulin}  \& {Yu}}{{Kaplinghat}
  et~al.}{2016}]{2016PhRvL.116d1302K}
{Kaplinghat} M.,  {Tulin} S.,   {Yu} H.-B.,  2016, \mn@doi [\prl]
  {10.1103/PhysRevLett.116.041302}, \href
  {https://ui.adsabs.harvard.edu/abs/2016PhRvL.116d1302K} {116, 041302}

\bibitem[\protect\citeauthoryear{{Kauffmann}, {White}  \&
  {Guiderdoni}}{{Kauffmann} et~al.}{1993}]{1993MNRAS.264..201K}
{Kauffmann} G.,  {White} S.~D.~M.,   {Guiderdoni} B.,  1993, \mn@doi [\mnras]
  {10.1093/mnras/264.1.201}, \href
  {https://ui.adsabs.harvard.edu/abs/1993MNRAS.264..201K} {264, 201}

\bibitem[\protect\citeauthoryear{{Keeton} \& {Moustakas}}{{Keeton} \&
  {Moustakas}}{2009}]{2009ApJ...699.1720K}
{Keeton} C.~R.,  {Moustakas} L.~A.,  2009, \mn@doi [\apj]
  {10.1088/0004-637X/699/2/1720}, \href
  {https://ui.adsabs.harvard.edu/abs/2009ApJ...699.1720K} {699, 1720}

\bibitem[\protect\citeauthoryear{{Kennedy}, {Frenk}, {Cole}  \&
  {Benson}}{{Kennedy} et~al.}{2014}]{2014MNRAS.442.2487K}
{Kennedy} R.,  {Frenk} C.,  {Cole} S.,   {Benson} A.,  2014, \mn@doi [\mnras]
  {10.1093/mnras/stu719}, \href
  {https://ui.adsabs.harvard.edu/abs/2014MNRAS.442.2487K} {442, 2487}

\bibitem[\protect\citeauthoryear{{Kim}, {Peter}  \& {Hargis}}{{Kim}
  et~al.}{2018}]{2018PhRvL.121u1302K}
{Kim} S.~Y.,  {Peter} A. H.~G.,   {Hargis} J.~R.,  2018, \mn@doi [\prl]
  {10.1103/PhysRevLett.121.211302}, \href
  {https://ui.adsabs.harvard.edu/abs/2018PhRvL.121u1302K} {121, 211302}

\bibitem[\protect\citeauthoryear{{King}}{{King}}{1962}]{1962AJ.....67..471K}
{King} I.,  1962, \mn@doi [\aj] {10.1086/108756}, \href
  {https://ui.adsabs.harvard.edu/abs/1962AJ.....67..471K} {67, 471}

\bibitem[\protect\citeauthoryear{{Klypin}, {Kravtsov}, {Valenzuela}  \&
  {Prada}}{{Klypin} et~al.}{1999}]{1999ApJ...522...82K}
{Klypin} A.,  {Kravtsov} A.~V.,  {Valenzuela} O.,   {Prada} F.,  1999, \mn@doi
  [\apj] {10.1086/307643}, \href
  {https://ui.adsabs.harvard.edu/abs/1999ApJ...522...82K} {522, 82}

\bibitem[\protect\citeauthoryear{{Kuzio de Naray} \& {Kaufmann}}{{Kuzio de
  Naray} \& {Kaufmann}}{2011}]{2011MNRAS.414.3617K}
{Kuzio de Naray} R.,  {Kaufmann} T.,  2011, \mn@doi [\mnras]
  {10.1111/j.1365-2966.2011.18656.x}, \href
  {https://ui.adsabs.harvard.edu/abs/2011MNRAS.414.3617K} {414, 3617}

\bibitem[\protect\citeauthoryear{{Kuzio de Naray} \& {Spekkens}}{{Kuzio de
  Naray} \& {Spekkens}}{2011}]{2011ApJ...741L..29K}
{Kuzio de Naray} R.,  {Spekkens} K.,  2011, \mn@doi [\apjl]
  {10.1088/2041-8205/741/2/L29}, \href
  {https://ui.adsabs.harvard.edu/abs/2011ApJ...741L..29K} {741, L29}

\bibitem[\protect\citeauthoryear{{Lacey} \& {Cole}}{{Lacey} \&
  {Cole}}{1993}]{1993MNRAS.262..627L}
{Lacey} C.,  {Cole} S.,  1993, \mn@doi [\mnras] {10.1093/mnras/262.3.627},
  \href {https://ui.adsabs.harvard.edu/abs/1993MNRAS.262..627L} {262, 627}

\bibitem[\protect\citeauthoryear{{Larson} et~al.,}{{Larson}
  et~al.}{2011}]{2011ApJS..192...16L}
{Larson} D.,  et~al., 2011, \mn@doi [\apjs] {10.1088/0067-0049/192/2/16}, \href
  {https://ui.adsabs.harvard.edu/abs/2011ApJS..192...16L} {192, 16}

\bibitem[\protect\citeauthoryear{{Lewin} \& {Smith}}{{Lewin} \&
  {Smith}}{1996}]{1996APh.....6...87L}
{Lewin} J.~D.,  {Smith} P.~F.,  1996, \mn@doi [Astroparticle Physics]
  {10.1016/S0927-6505(96)00047-3}, \href
  {https://ui.adsabs.harvard.edu/abs/1996APh.....6...87L} {6, 87}

\bibitem[\protect\citeauthoryear{{{\L}okas} \& {Mamon}}{{{\L}okas} \&
  {Mamon}}{2001}]{2001MNRAS.321..155L}
{{\L}okas} E.~L.,  {Mamon} G.~A.,  2001, \mn@doi [\mnras]
  {10.1046/j.1365-8711.2001.04007.x}, \href
  {https://ui.adsabs.harvard.edu/abs/2001MNRAS.321..155L} {321, 155}

\bibitem[\protect\citeauthoryear{{Lovell} et~al.,}{{Lovell}
  et~al.}{2012}]{2012MNRAS.420.2318L}
{Lovell} M.~R.,  et~al., 2012, \mn@doi [\mnras]
  {10.1111/j.1365-2966.2011.20200.x}, \href
  {https://ui.adsabs.harvard.edu/abs/2012MNRAS.420.2318L} {420, 2318}

\bibitem[\protect\citeauthoryear{{Lovell}, {Hellwing}, {Ludlow}, {Zavala},
  {Robertson}, {Fattahi}, {Frenk}  \& {Hardwick}}{{Lovell}
  et~al.}{2020}]{2020MNRAS.tmp.2452L}
{Lovell} M.~R.,  {Hellwing} W.,  {Ludlow} A.,  {Zavala} J.,  {Robertson} A.,
  {Fattahi} A.,  {Frenk} C.~S.,   {Hardwick} J.,  2020, \mn@doi [\mnras]
  {10.1093/mnras/staa2525}, \href
  {https://ui.adsabs.harvard.edu/abs/2020MNRAS.tmp.2452L} {}

\bibitem[\protect\citeauthoryear{{Lu}, {Benson}, {Mao}, {Tonnesen}, {Peter},
  {Wetzel}, {Boylan-Kolchin}  \& {Wechsler}}{{Lu}
  et~al.}{2016}]{2016ApJ...830...59L}
{Lu} Y.,  {Benson} A.,  {Mao} Y.-Y.,  {Tonnesen} S.,  {Peter} A. H.~G.,
  {Wetzel} A.~R.,  {Boylan-Kolchin} M.,   {Wechsler} R.~H.,  2016, \mn@doi
  [\apj] {10.3847/0004-637X/830/2/59}, \href
  {https://ui.adsabs.harvard.edu/abs/2016ApJ...830...59L} {830, 59}

\bibitem[\protect\citeauthoryear{{Maccio'}, {Sideris}, {Miranda}, {Moore}  \&
  {Jesseit}}{{Maccio'} et~al.}{2007}]{2007arXiv0704.3078M}
{Maccio'} A.~V.,  {Sideris} I.,  {Miranda} M.,  {Moore} B.,   {Jesseit} R.,
  2007, arXiv e-prints, \href
  {https://ui.adsabs.harvard.edu/abs/2007arXiv0704.3078M} {p. arXiv:0704.3078}

\bibitem[\protect\citeauthoryear{{Mao}, {Strigari}, {Wechsler}, {Wu}  \&
  {Hahn}}{{Mao} et~al.}{2013}]{2013ApJ...764...35M}
{Mao} Y.-Y.,  {Strigari} L.~E.,  {Wechsler} R.~H.,  {Wu} H.-Y.,   {Hahn} O.,
  2013, \mn@doi [\apj] {10.1088/0004-637X/764/1/35}, \href
  {https://ui.adsabs.harvard.edu/abs/2013ApJ...764...35M} {764, 35}

\bibitem[\protect\citeauthoryear{{Markevitch}, {Gonzalez}, {Clowe},
  {Vikhlinin}, {Forman}, {Jones}, {Murray}  \& {Tucker}}{{Markevitch}
  et~al.}{2004}]{2004ApJ...606..819M}
{Markevitch} M.,  {Gonzalez} A.~H.,  {Clowe} D.,  {Vikhlinin} A.,  {Forman} W.,
   {Jones} C.,  {Murray} S.,   {Tucker} W.,  2004, \mn@doi [\apj]
  {10.1086/383178}, \href
  {https://ui.adsabs.harvard.edu/abs/2004ApJ...606..819M} {606, 819}

\bibitem[\protect\citeauthoryear{{Markovi{\v{c}}} \& {Viel}}{{Markovi{\v{c}}}
  \& {Viel}}{2014}]{2014PASA...31....6M}
{Markovi{\v{c}}} K.,  {Viel} M.,  2014, \mn@doi [\pasa] {10.1017/pasa.2013.43},
  \href {https://ui.adsabs.harvard.edu/abs/2014PASA...31....6M} {31, e006}

\bibitem[\protect\citeauthoryear{{Mishra-Sharma}, {Van Tilburg}  \&
  {Weiner}}{{Mishra-Sharma} et~al.}{2020}]{2020PhRvD.102b3026M}
{Mishra-Sharma} S.,  {Van Tilburg} K.,   {Weiner} N.,  2020, \mn@doi [\prd]
  {10.1103/PhysRevD.102.023026}, \href
  {https://ui.adsabs.harvard.edu/abs/2020PhRvD.102b3026M} {102, 023026}

\bibitem[\protect\citeauthoryear{{Mondino}, {Taki}, {Van Tilburg}  \&
  {Weiner}}{{Mondino} et~al.}{2020}]{2020arXiv200201938M}
{Mondino} C.,  {Taki} A.-M.,  {Van Tilburg} K.,   {Weiner} N.,  2020, arXiv
  e-prints, \href {https://ui.adsabs.harvard.edu/abs/2020arXiv200201938M} {p.
  arXiv:2002.01938}

\bibitem[\protect\citeauthoryear{{Morningstar} et~al.,}{{Morningstar}
  et~al.}{2019}]{2019ApJ...883...14M}
{Morningstar} W.~R.,  et~al., 2019, \mn@doi [\apj] {10.3847/1538-4357/ab35d7},
  \href {https://ui.adsabs.harvard.edu/abs/2019ApJ...883...14M} {883, 14}

\bibitem[\protect\citeauthoryear{{Navarro}, {Frenk}  \& {White}}{{Navarro}
  et~al.}{1997}]{1997ApJ...490..493N}
{Navarro} J.~F.,  {Frenk} C.~S.,   {White} S. D.~M.,  1997, \mn@doi [\apj]
  {10.1086/304888}, \href
  {https://ui.adsabs.harvard.edu/abs/1997ApJ...490..493N} {490, 493}

\bibitem[\protect\citeauthoryear{{Newman}, {Treu}, {Ellis}, {Sand }, {Richard},
  {Marshall}, {Capak}  \& {Miyazaki}}{{Newman}
  et~al.}{2009}]{2009ApJ...706.1078N}
{Newman} A.~B.,  {Treu} T.,  {Ellis} R.~S.,  {Sand } D.~J.,  {Richard} J.,
  {Marshall} P.~J.,  {Capak} P.,   {Miyazaki} S.,  2009, \mn@doi [\apj]
  {10.1088/0004-637X/706/2/1078}, \href
  {https://ui.adsabs.harvard.edu/abs/2009ApJ...706.1078N} {706, 1078}

\bibitem[\protect\citeauthoryear{{Newman}, {Treu}, {Ellis}  \& {Sand
  }}{{Newman} et~al.}{2011}]{2011ApJ...728L..39N}
{Newman} A.~B.,  {Treu} T.,  {Ellis} R.~S.,   {Sand } D.~J.,  2011, \mn@doi
  [\apjl] {10.1088/2041-8205/728/2/L39}, \href
  {https://ui.adsabs.harvard.edu/abs/2011ApJ...728L..39N} {728, L39}

\bibitem[\protect\citeauthoryear{{Ngan} \& {Carlberg}}{{Ngan} \&
  {Carlberg}}{2014}]{2014ApJ...788..181N}
{Ngan} W.~H.~W.,  {Carlberg} R.~G.,  2014, \mn@doi [\apj]
  {10.1088/0004-637X/788/2/181}, \href
  {https://ui.adsabs.harvard.edu/abs/2014ApJ...788..181N} {788, 181}

\bibitem[\protect\citeauthoryear{{Ngan}, {Bozek}, {Carlberg}, {Wyse}, {Szalay}
  \& {Madau}}{{Ngan} et~al.}{2015}]{2015ApJ...803...75N}
{Ngan} W.,  {Bozek} B.,  {Carlberg} R.~G.,  {Wyse} R. F.~G.,  {Szalay} A.~S.,
  {Madau} P.,  2015, \mn@doi [\apj] {10.1088/0004-637X/803/2/75}, \href
  {https://ui.adsabs.harvard.edu/abs/2015ApJ...803...75N} {803, 75}

\bibitem[\protect\citeauthoryear{{Nierenberg} et~al.,}{{Nierenberg}
  et~al.}{2020}]{2020MNRAS.492.5314N}
{Nierenberg} A.~M.,  et~al., 2020, \mn@doi [\mnras] {10.1093/mnras/stz3588},
  \href {https://ui.adsabs.harvard.edu/abs/2020MNRAS.492.5314N} {492, 5314}

\bibitem[\protect\citeauthoryear{Parkinson, Cole  \& Helly}{Parkinson
  et~al.}{2008}]{parkinson_generating_2008}
Parkinson H.,  Cole S.,   Helly J.,  2008, Monthly Notices of the Royal
  Astronomical Society, 383, 557

\bibitem[\protect\citeauthoryear{{Pe{\~n}arrubia} \& {Benson}}{{Pe{\~n}arrubia}
  \& {Benson}}{2005}]{2005MNRAS.364..977P}
{Pe{\~n}arrubia} J.,  {Benson} A.~J.,  2005, \mn@doi [\mnras]
  {10.1111/j.1365-2966.2005.09633.x}, \href
  {https://ui.adsabs.harvard.edu/abs/2005MNRAS.364..977P} {364, 977}

\bibitem[\protect\citeauthoryear{{Peebles}}{{Peebles}}{1982}]{1982ApJ...263L...1P}
{Peebles} P.~J.~E.,  1982, \mn@doi [\apjl] {10.1086/183911}, \href
  {https://ui.adsabs.harvard.edu/abs/1982ApJ...263L...1P} {263, L1}

\bibitem[\protect\citeauthoryear{{Persic} \& {Salucci}}{{Persic} \&
  {Salucci}}{1988}]{1988MNRAS.234..131P}
{Persic} M.,  {Salucci} P.,  1988, \mn@doi [\mnras] {10.1093/mnras/234.1.131},
  \href {https://ui.adsabs.harvard.edu/abs/1988MNRAS.234..131P} {234, 131}

\bibitem[\protect\citeauthoryear{{Persic}, {Salucci}  \& {Stel}}{{Persic}
  et~al.}{1996}]{1996MNRAS.281...27P}
{Persic} M.,  {Salucci} P.,   {Stel} F.,  1996, \mn@doi [\mnras]
  {10.1093/mnras/278.1.27}, \href
  {https://ui.adsabs.harvard.edu/abs/1996MNRAS.281...27P} {281, 27}

\bibitem[\protect\citeauthoryear{{Peter} \& {Benson}}{{Peter} \&
  {Benson}}{2010}]{2010PhRvD..82l3521P}
{Peter} A. H.~G.,  {Benson} A.~J.,  2010, \mn@doi [\prd]
  {10.1103/PhysRevD.82.123521}, \href
  {https://ui.adsabs.harvard.edu/abs/2010PhRvD..82l3521P} {82, 123521}

\bibitem[\protect\citeauthoryear{{Planck Collaboration} et~al.,}{{Planck
  Collaboration} et~al.}{2011}]{2011A&A...536A...1P}
{Planck Collaboration} et~al., 2011, \mn@doi [\aap]
  {10.1051/0004-6361/201116464}, \href
  {https://ui.adsabs.harvard.edu/abs/2011A&A...536A...1P} {536, A1}

\bibitem[\protect\citeauthoryear{{Planck Collaboration} et~al.,}{{Planck
  Collaboration} et~al.}{2014a}]{2014A&A...571A...1P}
{Planck Collaboration} et~al., 2014a, \mn@doi [\aap]
  {10.1051/0004-6361/201321529}, \href
  {https://ui.adsabs.harvard.edu/abs/2014A&A...571A...1P} {571, A1}

\bibitem[\protect\citeauthoryear{{Planck Collaboration} et~al.,}{{Planck
  Collaboration} et~al.}{2014b}]{2014A&A...571A..16P}
{Planck Collaboration} et~al., 2014b, \mn@doi [\aap]
  {10.1051/0004-6361/201321591}, \href
  {https://ui.adsabs.harvard.edu/abs/2014A&A...571A..16P} {571, A16}

\bibitem[\protect\citeauthoryear{{Press} \& {Schechter}}{{Press} \&
  {Schechter}}{1974}]{1974ApJ...187..425P}
{Press} W.~H.,  {Schechter} P.,  1974, \mn@doi [\apj] {10.1086/152650}, \href
  {https://ui.adsabs.harvard.edu/abs/1974ApJ...187..425P} {187, 425}

\bibitem[\protect\citeauthoryear{{Pullen}, {Benson}  \& {Moustakas}}{{Pullen}
  et~al.}{2014}]{2014ApJ...792...24P}
{Pullen} A.~R.,  {Benson} A.~J.,   {Moustakas} L.~A.,  2014, \mn@doi [\apj]
  {10.1088/0004-637X/792/1/24}, \href
  {https://ui.adsabs.harvard.edu/abs/2014ApJ...792...24P} {792, 24}

\bibitem[\protect\citeauthoryear{{Randall}, {Markevitch}, {Clowe}, {Gonzalez}
  \& {Brada{\v{c}}}}{{Randall} et~al.}{2008}]{2008ApJ...679.1173R}
{Randall} S.~W.,  {Markevitch} M.,  {Clowe} D.,  {Gonzalez} A.~H.,
  {Brada{\v{c}}} M.,  2008, \mn@doi [\apj] {10.1086/587859}, \href
  {https://ui.adsabs.harvard.edu/abs/2008ApJ...679.1173R} {679, 1173}

\bibitem[\protect\citeauthoryear{{Ravi}, {Langellier}, {Phillips}, {Buschmann},
  {Safdi}  \& {Walsworth}}{{Ravi} et~al.}{2019}]{2019PhRvL.123i1101R}
{Ravi} A.,  {Langellier} N.,  {Phillips} D.~F.,  {Buschmann} M.,  {Safdi}
  B.~R.,   {Walsworth} R.~L.,  2019, \mn@doi [\prl]
  {10.1103/PhysRevLett.123.091101}, \href
  {https://ui.adsabs.harvard.edu/abs/2019PhRvL.123i1101R} {123, 091101}

\bibitem[\protect\citeauthoryear{{Relatores} et~al.,}{{Relatores}
  et~al.}{2019a}]{2019ApJ...873....5R}
{Relatores} N.~C.,  et~al., 2019a, \mn@doi [\apj] {10.3847/1538-4357/ab0382},
  \href {https://ui.adsabs.harvard.edu/abs/2019ApJ...873....5R} {873, 5}

\bibitem[\protect\citeauthoryear{{Relatores} et~al.,}{{Relatores}
  et~al.}{2019b}]{2019ApJ...887...94R}
{Relatores} N.~C.,  et~al., 2019b, \mn@doi [\apj] {10.3847/1538-4357/ab5305},
  \href {https://ui.adsabs.harvard.edu/abs/2019ApJ...887...94R} {887, 94}

\bibitem[\protect\citeauthoryear{{Robles} et~al.,}{{Robles}
  et~al.}{2017}]{robles2017}
{Robles} V.~H.,  et~al., 2017, \mn@doi [\mnras] {10.1093/mnras/stx2253}, \href
  {https://ui.adsabs.harvard.edu/abs/2017MNRAS.472.2945R} {472, 2945}

\bibitem[\protect\citeauthoryear{{Rubin}, {Ford}  \& {Thonnard}}{{Rubin}
  et~al.}{1980}]{1980ApJ...238..471R}
{Rubin} V.~C.,  {Ford} W.~K. J.,   {Thonnard} N.,  1980, \mn@doi [\apj]
  {10.1086/158003}, \href
  {https://ui.adsabs.harvard.edu/abs/1980ApJ...238..471R} {238, 471}

\bibitem[\protect\citeauthoryear{{Salucci}}{{Salucci}}{2001}]{2001MNRAS.320L...1S}
{Salucci} P.,  2001, \mn@doi [\mnras] {10.1046/j.1365-8711.2001.04076.x}, \href
  {https://ui.adsabs.harvard.edu/abs/2001MNRAS.320L...1S} {320, L1}

\bibitem[\protect\citeauthoryear{{Salucci}, {Wilkinson}, {Walker}, {Gilmore},
  {Grebel}, {Koch}, {Frigerio Martins}  \& {Wyse}}{{Salucci}
  et~al.}{2012}]{2012MNRAS.420.2034S}
{Salucci} P.,  {Wilkinson} M.~I.,  {Walker} M.~G.,  {Gilmore} G.~F.,  {Grebel}
  E.~K.,  {Koch} A.,  {Frigerio Martins} C.,   {Wyse} R. F.~G.,  2012, \mn@doi
  [\mnras] {10.1111/j.1365-2966.2011.20144.x}, \href
  {https://ui.adsabs.harvard.edu/abs/2012MNRAS.420.2034S} {420, 2034}

\bibitem[\protect\citeauthoryear{{Sembolini}, {Yepes}, {De Petris},
  {Gottl{\"o}ber}, {Lamagna}  \& {Comis}}{{Sembolini}
  et~al.}{2013}]{2013MNRAS.429..323S}
{Sembolini} F.,  {Yepes} G.,  {De Petris} M.,  {Gottl{\"o}ber} S.,  {Lamagna}
  L.,   {Comis} B.,  2013, \mn@doi [\mnras] {10.1093/mnras/sts339}, \href
  {https://ui.adsabs.harvard.edu/abs/2013MNRAS.429..323S} {429, 323}

\bibitem[\protect\citeauthoryear{{Simon}, {Bolatto}, {Leroy}, {Blitz}  \&
  {Gates}}{{Simon} et~al.}{2005}]{2005ApJ...621..757S}
{Simon} J.~D.,  {Bolatto} A.~D.,  {Leroy} A.,  {Blitz} L.,   {Gates} E.~L.,
  2005, \mn@doi [\apj] {10.1086/427684}, \href
  {https://ui.adsabs.harvard.edu/abs/2005ApJ...621..757S} {621, 757}

\bibitem[\protect\citeauthoryear{{Spingola}, {McKean}, {Auger}, {Fassnacht},
  {Koopmans}, {Lagattuta}  \& {Vegetti}}{{Spingola}
  et~al.}{2018}]{2018MNRAS.478.4816S}
{Spingola} C.,  {McKean} J.~P.,  {Auger} M.~W.,  {Fassnacht} C.~D.,  {Koopmans}
  L.~V.~E.,  {Lagattuta} D.~J.,   {Vegetti} S.,  2018, \mn@doi [\mnras]
  {10.1093/mnras/sty1326}, \href
  {https://ui.adsabs.harvard.edu/abs/2018MNRAS.478.4816S} {478, 4816}

\bibitem[\protect\citeauthoryear{{Springel} et~al.,}{{Springel}
  et~al.}{2008}]{2008MNRAS.391.1685S}
{Springel} V.,  et~al., 2008, \mn@doi [\mnras]
  {10.1111/j.1365-2966.2008.14066.x}, \href
  {https://ui.adsabs.harvard.edu/abs/2008MNRAS.391.1685S} {391, 1685}

\bibitem[\protect\citeauthoryear{{Taylor} \& {Babul}}{{Taylor} \&
  {Babul}}{2001}]{2001ApJ...559..716T}
{Taylor} J.~E.,  {Babul} A.,  2001, \mn@doi [\apj] {10.1086/322276}, \href
  {https://ui.adsabs.harvard.edu/abs/2001ApJ...559..716T} {559, 716}

\bibitem[\protect\citeauthoryear{{Van Tilburg}, {Taki}  \& {Weiner}}{{Van
  Tilburg} et~al.}{2018}]{2018JCAP...07..041V}
{Van Tilburg} K.,  {Taki} A.-M.,   {Weiner} N.,  2018, \mn@doi [\jcap]
  {10.1088/1475-7516/2018/07/041}, \href
  {https://ui.adsabs.harvard.edu/abs/2018JCAP...07..041V} {2018, 041}

\bibitem[\protect\citeauthoryear{{Vegetti} \& {Koopmans}}{{Vegetti} \&
  {Koopmans}}{2009}]{2009MNRAS.400.1583V}
{Vegetti} S.,  {Koopmans} L.~V.~E.,  2009, \mn@doi [\mnras]
  {10.1111/j.1365-2966.2009.15559.x}, \href
  {https://ui.adsabs.harvard.edu/abs/2009MNRAS.400.1583V} {400, 1583}

\bibitem[\protect\citeauthoryear{{Vegetti}, {Czoske}  \& {Koopmans}}{{Vegetti}
  et~al.}{2010}]{2010MNRAS.407..225V}
{Vegetti} S.,  {Czoske} O.,   {Koopmans} L. V.~E.,  2010, \mn@doi [\mnras]
  {10.1111/j.1365-2966.2010.16952.x}, \href
  {https://ui.adsabs.harvard.edu/abs/2010MNRAS.407..225V} {407, 225}

\bibitem[\protect\citeauthoryear{{Vegetti}, {Lagattuta}, {McKean}, {Auger},
  {Fassnacht}  \& {Koopmans}}{{Vegetti} et~al.}{2012}]{2012Natur.481..341V}
{Vegetti} S.,  {Lagattuta} D.~J.,  {McKean} J.~P.,  {Auger} M.~W.,  {Fassnacht}
  C.~D.,   {Koopmans} L.~V.~E.,  2012, \mn@doi [\nat] {10.1038/nature10669},
  \href {https://ui.adsabs.harvard.edu/abs/2012Natur.481..341V} {481, 341}

\bibitem[\protect\citeauthoryear{{Vegetti}, {Despali}, {Lovell}  \&
  {Enzi}}{{Vegetti} et~al.}{2018}]{2018MNRAS.481.3661V}
{Vegetti} S.,  {Despali} G.,  {Lovell} M.~R.,   {Enzi} W.,  2018, \mn@doi
  [\mnras] {10.1093/mnras/sty2393}, \href
  {https://ui.adsabs.harvard.edu/abs/2018MNRAS.481.3661V} {481, 3661}

\bibitem[\protect\citeauthoryear{{Viel}, {Bolton}  \& {Haehnelt}}{{Viel}
  et~al.}{2009}]{2009MNRAS.399L..39V}
{Viel} M.,  {Bolton} J.~S.,   {Haehnelt} M.~G.,  2009, \mn@doi [\mnras]
  {10.1111/j.1745-3933.2009.00720.x}, \href
  {https://ui.adsabs.harvard.edu/abs/2009MNRAS.399L..39V} {399, L39}

\bibitem[\protect\citeauthoryear{{Wang} \& {Zentner}}{{Wang} \&
  {Zentner}}{2012}]{2012PhRvD..85d3514W}
{Wang} M.-Y.,  {Zentner} A.~R.,  2012, \mn@doi [\prd]
  {10.1103/PhysRevD.85.043514}, \href
  {https://ui.adsabs.harvard.edu/abs/2012PhRvD..85d3514W} {85, 043514}

\bibitem[\protect\citeauthoryear{{Wang}, {Frenk}, {Navarro}, {Gao}  \&
  {Sawala}}{{Wang} et~al.}{2012}]{2012MNRAS.424.2715W}
{Wang} J.,  {Frenk} C.~S.,  {Navarro} J.~F.,  {Gao} L.,   {Sawala} T.,  2012,
  \mn@doi [\mnras] {10.1111/j.1365-2966.2012.21357.x}, \href
  {https://ui.adsabs.harvard.edu/abs/2012MNRAS.424.2715W} {424, 2715}

\bibitem[\protect\citeauthoryear{{Weinberg}}{{Weinberg}}{1986}]{1986ApJ...300...93W}
{Weinberg} M.~D.,  1986, \mn@doi [\apj] {10.1086/163785}, \href
  {https://ui.adsabs.harvard.edu/abs/1986ApJ...300...93W} {300, 93}

\bibitem[\protect\citeauthoryear{{Weinberg}}{{Weinberg}}{1994a}]{1994AJ....108.1398W}
{Weinberg} M.~D.,  1994a, \mn@doi [\aj] {10.1086/117161}, \href
  {https://ui.adsabs.harvard.edu/abs/1994AJ....108.1398W} {108, 1398}

\bibitem[\protect\citeauthoryear{{Weinberg}}{{Weinberg}}{1994b}]{1994AJ....108.1403W}
{Weinberg} M.~D.,  1994b, \mn@doi [\aj] {10.1086/117162}, \href
  {https://ui.adsabs.harvard.edu/abs/1994AJ....108.1403W} {108, 1403}

\bibitem[\protect\citeauthoryear{{Wolf} \& {Bullock}}{{Wolf} \&
  {Bullock}}{2012}]{2012arXiv1203.4240W}
{Wolf} J.,  {Bullock} J.~S.,  2012, arXiv e-prints, \href
  {https://ui.adsabs.harvard.edu/abs/2012arXiv1203.4240W} {p. arXiv:1203.4240}

\bibitem[\protect\citeauthoryear{{Yoon}, {Johnston}  \& {Hogg}}{{Yoon}
  et~al.}{2011}]{2011ApJ...731...58Y}
{Yoon} J.~H.,  {Johnston} K.~V.,   {Hogg} D.~W.,  2011, \mn@doi [\apj]
  {10.1088/0004-637X/731/1/58}, \href
  {https://ui.adsabs.harvard.edu/abs/2011ApJ...731...58Y} {731, 58}

\bibitem[\protect\citeauthoryear{{Zentner}, {Berlind}, {Bullock}, {Kravtsov}
  \& {Wechsler}}{{Zentner} et~al.}{2005}]{2005ApJ...624..505Z}
{Zentner} A.~R.,  {Berlind} A.~A.,  {Bullock} J.~S.,  {Kravtsov} A.~V.,
  {Wechsler} R.~H.,  2005, \mn@doi [\apj] {10.1086/428898}, \href
  {https://ui.adsabs.harvard.edu/abs/2005ApJ...624..505Z} {624, 505}

\bibitem[\protect\citeauthoryear{{de Blok}}{{de
  Blok}}{2010}]{2010AdAst2010E...5D}
{de Blok} W.~J.~G.,  2010, \mn@doi [Advances in Astronomy]
  {10.1155/2010/789293}, \href
  {https://ui.adsabs.harvard.edu/abs/2010AdAst2010E...5D} {2010, 789293}

\bibitem[\protect\citeauthoryear{{van den Bosch}, {Ogiya}, {Hahn}  \&
  {Burkert}}{{van den Bosch} et~al.}{2018}]{2018MNRAS.474.3043V}
{van den Bosch} F.~C.,  {Ogiya} G.,  {Hahn} O.,   {Burkert} A.,  2018, \mn@doi
  [\mnras] {10.1093/mnras/stx2956}, \href
  {https://ui.adsabs.harvard.edu/abs/2018MNRAS.474.3043V} {474, 3043}

\makeatother
\end{thebibliography}




\label{lastpage}
\end{document}